\documentclass[12pt,preprint]{aastex}

\slugcomment{Submitted to Astronomical Journal.}

\begin{document}


\title{B-Star Rotational Velocities in h and $\chi$ Persei: 
A Probe of Initial Conditions During the Star-Formation Epoch?}


\author{Stephen E. Strom\altaffilmark{1} and Sidney C. Wolff\altaffilmark{1}}
\affil{National Optical Astronomy Observatory, P.O. Box 26732,
Tucson, AZ 85726}
\email{sstrom@noao.edu}

\and
\author{David H. A. Dror\altaffilmark{1}}
\affil{Cornell University and National Optical Astronomy Observatory}


\altaffiltext{1}{NOAO is operated by the Association of Universities
for Research in Astronomy (AURA), Inc. under cooperative agreement 
with the National Science Foundation.}


\begin{abstract}
Projected rotational velocities (\textit{v sin i}) have been measured for 
216 B0--B9 stars in the rich, dense h and $\chi$ Persei double cluster 
and compared with the distribution of rotational velocities for a 
sample of field stars having comparable ages ($t \sim$ 12-15 Myr) and 
masses (M $\sim$ 4-15 $M\sun$). For stars that are relatively little evolved 
from their initial locations on the Zero Age Main Sequence 
(those with masses M $\sim$ 4-5 $M\sun$), the mean \textit{v sin i} measured 
for the h and $\chi$ Per sample is slightly more than 2 times larger 
than the mean determined for field stars of comparable mass, 
and the cluster and field \textit{v sin i} distributions differ with a 
high degree of significance.  For somewhat more evolved stars with 
masses in the range 5-9 $M\sun$, the mean \textit{v sin i} in h 
and $\chi$ Per is 
1.5 times that of the field; the \textit{v sin i} distributions differ as well, 
but with a lower degree of statistical significance.  For stars that 
have evolved significantly from the ZAMS and are approaching the 
hydrogen exhaustion phase (those with masses in the range  9-15 $M\sun$), the 
cluster and field star means and distributions are only slightly 
different. We argue that both the higher rotation rates and the pattern of 
rotation speeds as a function of mass that differentiate main sequence 
B stars in h and $\chi$ Per from their field analogs
were likely imprinted during the star formation process rather than 
a result of angular momentum evolution over the 12-15 Myr cluster lifetime. 
We speculate that these differences may reflect the effects of 
the higher accretion rates 
that theory suggests are characteristic of regions that give birth to 
dense clusters, namely: (a) higher initial rotation speeds; (b) higher
initial radii along the stellar birthline, resulting in greater 
spinup between the birthline and the ZAMS; and (c) a more 
pronounced maximum in the birthline radius-mass relationship
that results in differentially greater spinup for stars that become 
mid- to late- B stars on the ZAMS.
\end{abstract}



\keywords{stars: rotation---stars: 
formation---open clusters and associations:  
individual(\objectname{h and $\chi$ Persei})}


\section{Introduction}

Much of our current understanding of how stars form derives from the 
study of nearby star-forming regions such Taurus-Auriga, Ophiuchus, 
and Chamaeleon. These regions are populated by $\sim$ 100 young stars 
having typical masses $M < 1 M\sun$ contained within irregular molecular 
cloud complexes that span regions of size $\sim$ 3-10 pc.  However, 
the demographics and morphologies of these complexes differ markedly 
from those thought to produce the majority of stars over the history 
of the universe: dense stellar clusters containing $10^{4}$ to $10^{6}$ stars 
having masses ranging from 100 to 0.1 $M\sun$ formed within regions no 
more than $\sim$ 1pc in size. Do the dramatic differences in stellar density 
between these regions influence measurable properties of individual stars 
and the statistical properties (e.g. the initial mass function) of the 
emerging stellar populations? If so, what are the key physical causes? 
Answering these questions represents an essential first step toward 
developing a predictive theory of star formation of sufficient power 
to inform our understanding of how the mix of high and low mass stars 
populating galaxies  today came to be, and as a consequence how the 
observed relative abundances of heavy elements were established.

The importance and timeliness of these questions have stimulated several 
recent theoretical studies aimed at predicting initial protostellar 
conditions in dense star-forming complexes (e.g. McKee \& Tan 2003), and 
their relationship to emergent stellar mass functions (e.g. Elmegreen 
\& Shadmehri  2003). Dense stellar clusters form in regions of very 
high gas surface density characterized as well by close packing of 
protostars. The turbulent velocity of the gas in these regions is likely 
to be high, leading to (1) protostars of high initial density; 
(2) rapid protostellar collapse times; and (3) as a consequence, high 
time-averaged  accretion rates. The latter may be conducive to the 
formation of very high mass stars, since the dynamical pressure of accreting 
material can be high enough to overcome radiation pressure from the 
forming massive stars (McKee \& Tan 2003). The combination of protostellar 
cores characterized by higher turbulent speeds and higher mass accretion rates, 
combined with collisions between closely packed cores, 
may in turn produce a 'top-heavy' IMF (Elmegreen \& Shadmehri 2003).

These theoretical studies thus suggest: (1) higher time-averaged 
accretion rates; and (2) an initial mass function biased toward higher 
mass stars in high density stellar clusters. In principle, 
accretion rates characteristic of different star-forming regions can 
be diagnosed from the location of the stellar birthline (e.g. Stahler 1988) 
as determined from spectroscopic and photometric observations of 
precision sufficient to locate pre-main sequence stars spanning a 
range of masses in the HR diagram, provided that the target regions 
are young and accurate age estimates are available. Quantifying initial 
mass functions requires similar observations.

To date, it has not been possible to determine either birthline locations 
or IMFs spanning the full range of stellar masses primarily because 
the best and closest examples of high density clusters 
(e.g. Arches at the Galactic Center, Stolte et al. 2002; R 136 in 
the Magellanic  Clouds, Massey \& Hunter 1998; Sirianni et al. 2000) 
suffer from extreme crowding, which thus far has limited 
photometric and spectroscopic observations to main-sequence stars and a 
few pre-main-sequence stars with masses above $\sim2 M\sun$. Such stars are 
both bright enough to stand out relative to the stellar background and 
rare enough to avoid overlap with objects of comparable brightness.  
Next generation adaptive optics systems on current generation large 
telescopes have the potential to overcome the limitations of crowding 
and enable determination of stellar luminosities and 
effective temperatures for stars with masses as small as 0.1 $M\sun$, a 
level more than sufficient to confront theoretical predictions of IMF 
shape and birthline positions. However, until such systems become 
operational, other and less direct approaches must suffice. We explore here the 
possibility that the distribution of stellar rotational velocities 
can provide a surrogate indicator of the differences in initial conditions 
between low and high density star-forming regions. Our reasoning is as follows. 

Current theory suggests that initial stellar angular momenta are 
established during the primary stellar accretion phase via locking of 
stellar angular velocity to the angular velocity of the 
circumstellar accretion disk at or near the radius, r(m), where the 
stellar magnetosphere links to the disk (e.g. K\"{o}nigl 1991; 
Shu et al. 1994). That radius is set by the balance between the 
dynamical pressure of accreting material and the magnetic pressure of 
the magnetic field rooted in the forming star. For a fixed stellar 
magnetic field strength, the higher the accretion rate 
through the disk, the smaller r(m), the higher the Keplerian rotation 
speed of the disk at r(m), and hence the higher the angular rotational 
speed of the star.  Hence, if accretion rates are higher in high 
density star-forming regions, 
the resulting stellar population would be expected to exhibit 
higher rotation speeds on average.

Supporting observational evidence in the literature is sparse, but 
highly suggestive. For example, Wolff, Edwards, \& Preston (1982;
hereafter WEP) note 
that the distribution of rotation speeds among B stars in 
the relatively dense Orion Nebula Cluster is significantly shifted toward 
higher values as compared to stars of similar type distributed among 
the much lower density regions of the Orion star-forming complex. 
Similarly, Slettebak (1968) argues that stellar rotation speeds among the 
luminous B giant stars located in the vicinity of the extremely dense 
h and $\chi$ Persei double cluster are $\sim$ 50\% higher than their field 
counterparts. Moreover, he reports an unusually high number of rapidly rotating Be stars, 
possibly indicative of a higher fraction of rapidly rotating 
stars in h and $\chi$ Per. However, the Orion study includes only a 
modest sample of stars, while past discussions of h and $\chi$ Per rotation 
properties suffer from concerns regarding  the similarity in age range 
among field and cluster cohorts combined with  the possibility of 
evolutionary changes in stellar angular momenta. Confronting the hypothesis 
that stars born in dense clusters rotate more rapidly than their 
field counterparts requires sufficiently large samples 
of cluster and field stars spanning an identical range of ages.

The goal of this contribution is to effect a robust statistical comparison 
of the distribution of stellar rotational velocities for a sample of 
B0-B9 stars in h and $\chi$ Persei (typical stellar density 
of $10^{4}$ pc$^{-3}$; Slesnick, Hillenbrand, \& Massey, 2002, 
hereafter SHM) with those observed among field B stars of 
comparable age as determined from 
their location in the Str\"{o}mgren  ($\beta$, c$_{0}$) plane. 
This latter sample is almost certainly dominated by stars born in 
much lower density environments.

\section{New Observations of B Stars in h and $\chi$ Persei}
\subsection{Spectroscopy}

We report here new rotational velocity determinations of 216 stars in 
h and $\chi$ Per with estimated masses greater than 4 $M\sun$. These stars 
were chosen from the recent photometric and moderate resolution spectroscopic 
study of the double cluster by SHM. The basic data for the h and $\chi$ Per 
stars are listed in Table 1.  Column 1 provides an identification from 
SHM; column 2 lists the Oosterhoff number; columns 3 and 4 list respectively 
the absolute visual magnitude M$_{V}$ and the log of the effective temperature 
derived by SHM; column 5 lists the spectral type if available;  column 6 
lists the mass derived by SHM; column 7 lists the derived value 
of \textit{v sin i} 
(see below); column 8 lists a group assigned to the star on the basis of 
its effective temperature(see below); in columns 9 and 10 we list the 
Str\"{o}mgren indices c$_{0}$ and $\beta$, respectively, from the work of 
Capilla \& Fabregat (2002) or Crawford, Glaspey, \& Perry, (1970); and 
in column 11, we indicate whether the star is judged to be a spectroscopic 
binary with velocity amplitude K \texttt{>} 30 km/sec (see below). The HR diagram 
for these stars is shown in Figure 1.  Also shown are evolutionary tracks 
from Schaller et al. (1992), with the M$_{bol}$ from these tracks converted 
to M$_{V}$ by using the relationship between bolometric correction and 
T$_{eff}$ given by SHM.  

For the purpose of later analysis, we divide the h and $\chi$ Per stars 
into three temperature groups, which are shown in Figure 1.  We 
effected this division in order to examine separately the rotational 
properties of: (1) relatively unevolved stars, still located within 
0.5 mag of the ZAMS; (2) stars located within 1 mag of the ZAMS; and 
(3) stars that have  evolved significantly from the ZAMS.  Comparison 
with the evolutionary tracks indicates that these three groups 
correspond to mass ranges of, approximately, 3.5-5 $M\sun$, 5-9 $M\sun$, 
and 9-15 $M\sun$.  These groups are identified in Table 1 (column 8) as 
Group 1 (coolest and close to the ZAMS), Group 2 (middle range of 
temperature and slightly evolved from the ZAMS), and Group 3 (hottest and 
most evolved), respectively.
  
Spectra of the h and $\chi$ Per stars in our sample were obtained 
during three nights in September, 2002, with the Hydra multi-fiber 
spectrograph and the WIYN 3.5-meter telescope on Kitt Peak. 
The 316/63.4 echelle grating and a narrow-band order-separating filter 
were used in conjunction with the red bench camera to produce spectra 
with resolution $\sim$0.2 \AA\ centered at a wavelength of 
$\lambda$4461 \AA\ and spanning 120 \AA. This wavelength 
region was selected in order to 
include the two strong features He I $\lambda$4471 and 
Mg II $\lambda$4481, which together provide the basis for 
determining accurate rotational velocities for stars in the desired spectral 
type range: B0-B9.  

Eight separate fiber settings enabled observations of a total of 
216 stars. Three settings were targeted at the brighter members of 
the cluster (8.5 \texttt{<} B \texttt{<} 12) and five at the fainter
members (12 \texttt{<} B \texttt{<} 14.5). 
Exposure times were 30-60 minutes (divided among three exposures) for the 
bright sample and 90-120 minutes (divided among 3-4 exposures) for the 
faint sample. The series of three exposures for the bright stars was 
repeated either one or two nights later.  Flat-field 
exposures and wavelength calibration observations derived from Th-Ar 
lamp spectra were obtained before or after each exposure.

The resulting spectra were extracted, cosmic-ray cleaned, combined and 
wavelength-calibrated using standard DOHYDRA IRAF reduction scripts. 
The resulting values of signal/noise ranged from 25 to 100 per 
resolution element for a typical target.

We also obtained spectra of 25  B1-B9 stars in the I Lac association; 
18 of these stars, with spectral types in the range B1-B3, have published 
projected rotational velocities spanning the range 20-365 km s$^{-1}$ 
(Abt \& Hunter 1962) and served as standards.

\subsection{Derived Rotational Velocities} 

Rotational velocities for the hotter stars in our sample (stars in 
the two higher temperature groups shown in Figure 1) were determined 
(1) by establishing the relationship between the full-width at half maximum 
(FWHM) for He I and Mg II and projected rotational velocity (\textit{v sin i}) 
for the I Lac rotational standard stars; and (2) using this relationship
and the measured FWHM to establish \textit{v sin i} for the
unknowns (following 
Abt, Levato, \& Grosso 2002, hereafter ALG). FWHM was determined from
a Gaussian fit to the observed profiles. For the stars in 
these two groups, the He I line is substantially stronger than the 
Mg II line and was given twice the weight in deriving the 
average \textit{v sin i} 
from the two lines.  The relationships between FWHM and \textit{v sin i} 
for the 
standard stars are shown in Figure 2, which demonstrates a good correlation 
between line width and \textit{v sin i}. 

Several stars in our h and $\chi$ Per sample that fall in our two higher 
temperature groups have also been observed by Gies \& Huang (2003 and 
private communication).  These authors derive values of \textit{v sin i} 
by fitting 
three He I lines ($\lambda\lambda$4026, 4387, and 4471) with profiles 
derived from model atmospheres. Their computed profiles take limb and 
gravity darkening into account.  Figure 3 shows a comparison between 
the values of \textit{v sin i} derived by Gies \& Huang and in the 
current study.  
The best fitting straight line is given by
\begin{equation}
	v sin i (Gies \& Huang) = 1.05(\pm0.06) v sin i 
	(current study) + 11 (\pm9) km s^{-1}.
\end{equation}
Our results thus correlate well with the measurements of Gies \& Huang
but are systematically smaller by about 5\%.  Gies \& Huang in turn state 
that their calibration agrees with the calibrations of Slettebak (1968; 
1985) with a best fit slope of 0.999 and a zero point offset of 34 
km s$^{-1}$ in the sense that the Gies \& Huang measurements are smaller 
than the Slettebak values.  We have 13 stars in common with Slettebak and 
find that $\textit{v sin i} (Slettebak) = 1.13 \textit{v sin i} 
(current study) + 39 km s^{-1}$, 
again indicating that our results are systematically slightly 
smaller. The Slettebak measurements were made from photographic plates
with dispersions of 40 and 47 \AA\ mm$^{-1}$ and were insensitive to 
rotations less than about 50 km s$^{-1}$, which fact likely accounts for 
the zero-point offset.
 
To calibrate the rotational velocities for stars in the coolest 
temperature group, we made use of a previous set of Hydra observations
of stars with low rotational velocities measured by ALG and 
having spectral types in the range B0-B8 (HR 1855, B0V, \textit{v sin i} 
= 10 km/sec; 
HR 2222, B1V, \textit{v sin i} = 0 km/sec; HR 153, B2IV, 
\textit{v sin i} = 10 km/sec; 
HR 6042, B5V, \textit{v sin i} = 30 km/sec; and HR 677, B8V, 25 km/sec).  We 
artificially broadened these spectra by convolving the observed 
standard star spectra with a rotational-broadening profile corresponding 
to projected velocities of 50, 100, 150, 200, 300, 350, and 400 km/s. 
Over this entire spectral range, we find that the FWHMs of the calculated
broadened line profiles are well correlated with rotation speed and
that the slope and zero point of the relationships for $\lambda$4471 and
$\lambda$4481 do not vary significantly with spectral type.  
Guided by this result, we chose 
to adopt the (\textit{v sin i}, FWHM) relationships derived 
empirically for the two 
hotter groups for the cooler group as well. Because the He I and 
Mg II lines have similar equivalent widths for stars in the coolest group, 
the two lines were given equal weight in deriving \textit{v sin i}.

Line widths become relatively insensitive to rotation once the rotation 
rate approaches the critical velocity owing to the effects of gravity 
darkening.  Townsend, Owocki, \& Howarth (2004) have constructed models of 
rapidly rotating B stars and find that for B0-B7 stars viewed equator-on 
and rotating at 95\% of the critical velocity, the measured velocity 
will be up to 17\% too low for measurements of Mg II 4481 and up to 33\% 
too low for measurements of He I 4471 if gravity darkening is ignored. The 
effects are much smaller both for lower rotation rates and 
smaller angles of inclination.  For stars in our sample, critical 
velocities range from about 400 to more than 500 km/sec.  We find that 
N(\textit{v sin i}), the distribution of apparent \textit{v sin i}, decreases 
rapidly with increasing rotation above 250 km/sec.  Only 
about 12\% of the h and $\chi$ Per stars in the two lowest mass bins, and 
fewer than 5\% of the field stars, appear to rotate faster 
than 300 km/sec (cf. Fig. 6).  Hence, the fraction of our sample that 
might be strongly affected by gravity darkening is 
insignificant (\texttt{<} 10\%).   

The values of \textit{v sin i} derived from the FWHMs are listed in column 7 of 
Table 1.  Where more than one observation of a star is available, the 
quoted \textit{v sin i} value represents an average of all determinations.  The 
internal accuracy of our \textit{v sin i} determinations can be assessed 
by comparing 
estimates derived from independent observations obtained on different nights. 
From such a comparison, we conclude that our reported \textit{v sin i} 
values have 
an internal uncertainty of $\sim$10\%.  From the comparisons with the data of
Gies \& Huang and of Slettebak for h and $\chi$ Per, we 
have shown that the data transform to an externally calibrated system with 
a systematic uncertainty of about 5-10\%.

\subsection{Radial Velocities and the Search for Binaries}

A number of studies (e.g. Abt \& Hunter 1962; ALG) suggest possible 
correlations between binarity and observed rotational velocity. We have 
two methods for detecting binaries with our data set:  1) we can identify 
those spectra that have double lines; and 2) we can look for stars 
with radial velocities that differ significantly from the cluster mean.  
For each star in our sample, we have derived a radial velocity from the 
observed line centroid wavelengths of the He I and Mg II lines. 

The internal accuracy of the velocity determinations was judged by 
comparing the mean velocity derived from the He I and Mg II lines for 
three individual 1800 sec exposures that when summed 
constituted the 120 min observation of one of the faint fields. Because 
the pairs of exposures are separated by no more than 45 min, we expect any 
intrinsic radial velocity variations over this short time to be negligible 
compared to measurement errors.  This comparison should give us a 
worst case estimate of the errors since the S/N of the observations of 
the faint stars is somewhat lower than for the bright stars and because 
(as we show in Section 4) the rotation rates of the fainter stars are 
somewhat higher and broader lines are harder to measure.

In Figure 4, we depict the cumulative distribution of velocity differences 
for one pair of exposures for the fainter stars in our sample. Note that
about 85\% of the stars have velocity differences that are less than 10 km/s. 
Only about 3\% of the stars have velocity differences 
greater than 30 km/s, and the fact that a few stars have large errors 
is a consequence of the difficulty of measuring centroids of very broad lines.

This result provides the basis for compiling a list of candidate 
spectroscopic binaries. To be considered a candidate binary, the average 
velocity of a star on the summed exposures had to differ from the mean 
cluster velocity derived from the full sample of 216 stars by 30 km/sec.  In 
addition, a few stars showed double lines.  Column 11 of Table 1 
indicates which stars are candidate spectroscopic binaries and the 
reason for their candidacy.  For double-lined stars, the 
velocity listed is the difference in velocity of the two components.  
For single-lined stars, the table gives the difference between the 
stellar velocity and the cluster mean.  The velocities 
marked with a colon are based on a single line. We have also compared the 
differences in velocity for the observations of bright stars taken on 
two different nights. Star 33 shows a change in velocity 
of 200 km/sec, which confirms its binary nature. Star 150 shows a 
velocity difference of 34 km/sec, just barely significant  given our 
criteria. This star may also be a binary but has not been 
so designated in Table 1. We note finally, that our criterion
for selecting candidate spectroscopic binaries could result
in missing objects with velocity differences relative to
the cluster mean close to 30 km/s in
cases where the rotation speed of the primary exceeds 300 km/s.
However, as noted previously, such 
rapidly rotating objects comprise less than 10 \% of our 
field and cluster samples.

\section{The Comparison Sample: Field B Stars}

The B stars in h and $\chi$ Per are members of high stellar density 
bound systems. Isolated field B stars of ages comparable
to that of h and $\chi$ Per ($\sim$12 Myr) are most likely drawn from stars 
born in much lower density environments: (a)  stars formed initially 
in isolation or in small aggregates; or (b) stars whose peculiar 
motions have carried them several tens of parsecs 
away from their birthplaces in loose OB associations. 
The population of isolated field B stars may also contain 
a small number of stars born in dense environments but later ejected via
gravitational encounters (runaway stars).

Extensive observations of rotational velocities for bright field B stars 
are available in the literature. The recent study by ALG provides a homogeneous 
database for a large (1092 stars) sample of field B stars listed in the Bright Star Catalog.  
Their observations were obtained with a CCD and have a resolution of 
0.11 A or 7.1 km/sec.  The rotational velocities quoted by ALG were 
calibrated against Slettebak et al. (1975) standards.  Since Gies \& Huang 
have shown that their data for h and $\chi$ Per are consistent with this 
system, and our measurements are about 5\% smaller than the Gies \& Huang
values, a comparison between our h and $\chi$ Per data and the ALG data 
should be valid to within the externally-calibrated uncertainties 
of $\sim$5-10\%. We note that while the ALG sample, drawn from the 
Bright Star Catalog, is dominated by isolated field stars, it contains
as well a very modest number of stars located in relatively dense 
environments (e.g. the Orion Nebula Cluster). We have not attempted
to exclude such stars from the sample, but note that including them
will tend to reduce any
differences between the distribution of rotational velocities between
our h and $\chi$ Per and field samples.

The surface rotation rates of stars can be expected to change as stars 
evolve because of the changing  moment of inertia of the star; transport 
of angular momentum within the star; and possible loss of angular momentum 
due to winds. Hence, in order to assess intrinsic differences 
in the distribution of rotation speeds among stars in h and $\chi$ Per 
and the field, it is essential that the field star sample include only
objects having ages comparable to h and $\chi$ Per (12-15 Myr; see Figure 1 
and Slesnick et al. 2002).  To do this requires luminosity and effective 
temperature values of precision sufficient to enable age estimates. Because
most stars in the ALG catalog lack parallaxes accurate enough to derive 
luminosities relative to the Zero Age Main Sequence (ZAMS) and thus 
stellar age, we have established their evolutionary state by 
using the Str\"{o}mgren 
$\beta$ and c$_{0}$ indices. These indices provide accurate estimates 
of surface 
gravity and effective temperature respectively, thus allowing evaluation 
of stellar ages:  the youngest stars will have the highest surface 
gravities for fixed effective temperature (large $\beta$ index at constant 
c$_{0}$), whereas more evolved stars will have lower surface gravities and thus 
smaller $\beta$ indices. 

In order to select field stars of ages comparable to those of the 
h and $\chi$ Per sample, we made use of the 
Str\"{o}mgren $\beta$ and c$_{1}$ indices listed by Hauck \& Mermilliod (1998) for the 
ALG sample. For h and $\chi$ Per, values of $\beta$ and c$_{1}$ are available 
from measurements reported by Crawford, Glaspey, \& Perry (1970) and by 
Capilla \& Fabregat (2002). Apropos photometry of the ALG sample,
Hauck \& Mermilliod (1998) have carefully transformed heterogeneous
data from the literature to the Crawford system, noted 
discrepant values, and given them low weight. For h and $\chi$ Per, 
Capilla \& Fabregat (2002) have used as standard stars objects in
clusters measured by Crawford and collaborators, or objects measured
by other investigators who made use of the same photometer-telescope
combinations as Crawford. We thus believe that the Str\"{o}mgren
photometry for both the ALG field stars and the h and $\chi$ Per sample 
has been transformed carefully to the Crawford system and can thus be intercompared
with confidence.

For the field stars, we used the relationship between intrinsic 
color (B-V) and spectral type (Drilling \& Landolt 2000) to estimate 
the reddening and calculate  the reddening-corrected  Str\"{o}mgren index, 
c$_{0}$. For those few stars for which values of B-V were not available 
from the Simbad database, we have assumed zero reddening.  Given that the 
reddening E(c$_{1}$) for the B stars for which we do have color information 
is typically 0.01-0.02 mag, and seldom exceeds 0.03 mag., any uncertainties 
in the reddening are unimportant for the analysis in this paper.  
The c$_{1}$ indices observed for the h and $\chi$ Per sample have been 
transformed to c$_{0}$ by assuming an average reddening of $E(b-y) = 0.4$
(i.e. $E(c_{1}) = 0.2E(b-y)$;  see Capilla \& Fabregat 2002). The relationships
between $\beta$ and M$_{V}$ and c$_{0}$ and T$_{eff}$ coupled with the
group boundaries shown in Figure 1 were then used to 
establish boundaries between groups 1, 2 and 3 in the ($\beta$,c$_{0}$) plane.
The subset of the ALG sample falling within these boundaries is plotted in Figure 
5; the symbols indicate those stars in the ALG sample that correspond to 
Groups 1, 2 and 3. Also shown in this Figure are the location of the subset
of stars in the h and $\chi$ Per sample for which published Str\"{o}mgren
photometry is available.

Table 2 lists the values of $\beta$ and c$_{0}$ (or c$_{1}$ for the stars 
lacking B-V measurements; see above) for the field stars plotted in 
Figure 5 along with the values of \textit{v sin i} from ALG and the 
assignment of the star to one of the three temperature groups 
defined for the h and $\chi$ Per sample.

We note that the field stars in the ALG sample that fall within 
low (Group 1) and intermediate (Group 2) temperature
groups comprise not only objects
of luminosity class V but of luminosity classes III and even II
(though the latter comprise $ < 1\% $ of Groups 1 and 2).
At first glance, this would appear surprising, given that stars
in Groups 1 and 2 are expected to be little evolved from their
initial location on the ZAMS and thus to have reported spectra consistent
with assignment to luminosity class V. However, 
in the temperature range spanned by these two groups, the
actual difference in luminosity between class III and
class V is quite modest. Quantitatively, the difference
in the $\beta$,c$_{0}$ plane between the mean relationships for 
luminosity classes V and III is 0.04 mag in $\beta$, which corresponds to
only 0.4 mag in absolute visual magnitude (Crawford, 1978). By comparison,
the full range of $\beta$ values at a given c$_{0}$  among late B
stars nominally assigned to luminosity class V is almost 0.1 mag,
thus resulting in significant overlap in $\beta$ values with stars
assigned to luminosity class III. Hence, the appearance of some stars
classified as luminosity class III among the objects assigned to
Groups 1 and 2 on the basis of their location in the ($\beta$,c$_{0}$) plane
is in fact expected. That the range of $\beta$ values is as large as 0.1 mag
for a given spectral type and luminosity class, almost certainly reflects both
the subjectivity inherent in any visual classification, as well as small
errors introduced by assigning discrete spectral types as opposed
to a continuously varying indicator of effective temperature, c$_{0}$.

Because ($\beta$,c$_{0}$) photometry 
provides finer resolution in both temperature and luminosity
than MK classification, and because we have
confidence that the ALG and h and $\chi$ Per sample have been
transformed appropriately to the Crawford system, we believe that
using the observed locations of the ALG and h and $\chi$ Per stars
in the ($\beta$,c$_{0}$) plane provides the most reliable means of
sorting each sample into identical temperature and age groups.

Could the use of photometric indices as opposed to
spectroscopic sorting introduce any subtle selection biases?
One concern is that rapid rotation may alter observed $\beta$ and 
c$_{0}$ indices sufficiently to either exclude rapidly rotating stars from the sample, 
or to move them across the boundaries defining our three
temperature groups. Empirical (Crawford, 1978) and theoretical
(Collins \& Sonneborn, 1977) studies suggest that changes
in $\beta$ and c$_{0}$, driven by temperature and gravity
variations as a function of latitude among rotationally
distorted stellar surfaces, are in practice significant only for 
stars having rotation speeds
in excess of 250 km/sec. The most rapidly rotating stars among
the cohort with rotation speeds in excess of 250 km/sec may exhibit
H$\beta$ emission, the presence of which could change
measured $\beta$ sufficiently to drive the star outside the
bins used to define our groups. However, examination of the ALG sample suggest that
only $8 \%, 6 \%$, and $3 \%$ of stars in the temperature range
spanned by groups 1, 2 and 3 respectively have rotation speeds
higher than 250 km/sec. These percentages represent strong upper limits
on the actual fraction of stars that would either be excluded 
from our sample completely, or moved from one group to another.
Consequently, we believe that selection via location in
the $\beta$,c$_{0}$ plane will not produce significant biases
in the derived distributions of rotation speeds.

For the purpose of assessing whether the 
frequency of close binaries has an effect on 
the comparison of rotational velocities of the h 
and $\chi$ Per sample with stars in the field, we have searched the 
9$^{th}$ spectroscopic binary catalog (http://sb9.astro.ulb.ac.be) for
orbital parameters for stars in the ALG sample.  In h and $\chi$ Per, we 
can detect binaries only if they differ from the cluster mean velocity
by more than 30 km s$^{-1}$ or if we see double lines. Therefore, 
in Table 3, we list 
those field stars that (a) meet our color criteria; (b) have 
orbital semi-amplitudes greater than 30 km s$^{-1}$; and/or (c) for which 
ALG reported seeing double lines.  We recognize that the colors of 
spectroscopic binaries do not provide an accurate reflection of the 
temperature and surface gravity of the primary star, but we cannot make 
corrections for this effect for either h and $\chi$ Per or the field stars, 
and so will treat both samples identically.

\section{Distribution of Rotational Velocities}

In Figure 6, we plot the frequency distributions of rotational 
velocity N(\textit{v sin i}) for each of the three groups we have defined.  
All of 
the stars in our samples are included in the plots; these include the 
spectroscopic binary candidates identified in h and $\chi$ Per as well 
as the primaries of binaries in the field star sample. We have excluded 
the secondaries in the field star sample because in most cases we lack 
the temperature data needed to assign them to one of our three groups. 
In Figure 7, we plot the cumulative distributions for the \textit{v sin i} data 
shown in Figure 6.  These figures suggest that the N(\textit{v sin i}) 
distribution for the h and $\chi$ Per group 1 stars (those that are little
evolved from the ZAMS) is strikingly different 
from that of the field stars of similar mass and age.
The distribution of \textit{v sin i} for the middle group of stars in h 
and $\chi$ Per also differs from that of the field stars in the same 
temperature range, but the difference is smaller.  The distributions of 
\textit{v sin i} for evolved stars in the hottest group (group 3) 
exhibit only a small difference.  

In all three temperature ranges, the sense of the difference is the same: 
while the number of rapid rotators (\textit{v sin i} 
\texttt{>} 250 km/sec) and 
the maximum 
rotation velocity measured are similar for both field and cluster stars, 
there is a marked deficiency of slow rotators 
(\textit{v sin i} \texttt{<} 100 km/sec) among 
the cluster stars. In order to assess the statistical significance of 
these differences, we calculated the corresponding KS probabilities that 
the distributions are drawn from the same distribution. These 
probabilities are 2x10$^{-10}$, 9x10$^{-6}$, and 9x10$^{-3}$ for 
regions (1), (2), and (3) respectively. 

Table 4 summarizes the average rotational velocities 
\texttt{<}\textit{v sin i}\texttt{>} for the 
stars in h and $\chi$ Per and the field.  If we consider all of the stars 
in the sample, including binaries, we find that for the unevolved stars 
in coolest temperature region (Group 1), the mean \textit{v sin i} 
in h and $\chi$ Per is twice 
that of their field counterparts and that the distributions are 
different with a high degree of significance. For stars in middle 
temperature range, the \texttt{<}\textit{v sin i}\texttt{>} 
in h and $\chi$ Per is 1.56 times that of 
the field, and the distributions are different, with a lower but still 
high degree of significance. For the evolved stars in region (3), 
\texttt{<}\textit{v sin i}\texttt{>} 
in h and $\chi$ Per is only 1.25 times that of the field, and the 
distributions differ, but only marginally.

As a check on the robustness of our result, we computed mean values
of vsini for the field stars using published spectral types and
luminosity classes as opposed to Str\"{o}mgren photometry. For the
purposes of this test, the 
temperature boundaries for groups 1, 2 and 3 were defined to be
B5-B9, B2.5-B5 and B0-B2 respectively. Stars with luminosity
classes III-V were included. The resulting mean values are 112 km/s,
109 km/s, and 113 km/s respectively for groups 1,2 and 3. While
these values differ by ~15-20\% from those listed in Table 4,
our basic conclusions regarding the differences between the
h and $\chi$ Per and field samples remain the same: the rotation
speeds for groups 1 and 2 are significantly higher in h and $\chi$ Per
as compared with the field, while for group 3, the rotation speeds
between cluster and field are indistinguishable statistically.

The number of detectable spectroscopic binaries in h and $\chi$ Per is 
insufficient to enable comparison of the \textit{distribution} of rotational 
velocities for the binaries among the three separate groups at a high 
level of statistical significance.  Instead, we have calculated the mean 
rotational speeds, \texttt{<}\textit{v sin i}\texttt{>}, 
for the binaries in each of the groups.  We 
note that assignments to each of the groups are based on observed colors, 
which reflect the luminosity-weighted contributions from primary and secondary 
components. For all binaries the effective temperature of the primary 
assigned on the basis of color will thus be smaller than its true 
effective temperature. 

The values of \texttt{<}\textit{v sin i}\texttt{>} for the 
binaries with K \texttt{>} 30 km/sec and for 
the complementary samples excluding  the binaries are compared in Table 4.  
For group 1, the field binaries have a 
\texttt{<}\textit{v sin i}\texttt{>} value 
only 58\% as large as
\texttt{<}\textit{v sin i}\texttt{>} for the field stars not in known 
binaries with K \texttt{>} 30 km/sec.  
For groups 1 and 2,  the differences in 
\texttt{<}\textit{v sin i}\texttt{>} between the binaries
with K \texttt{>} 30 km/sec and the remaining stars in the same temperature 
range are not significant.  Because only $\sim$10\% of the stars show evidence 
of radial velocity variations greater than 30 km/sec, we conclude that the 
distributions of rotation speeds for h and $\chi$ Per and the field 
presented above are not affected significantly by the inclusion or 
exclusion of binaries from the sample.

\section{Discussion}

Our results show that relatively unevolved stars in h and $\chi$ Per 
(those in group 1 that presumably reflect their initial angular momenta 
most accurately) rotate on average more rapidly than stars of comparable 
age in the field by about a factor of 2.  The stars with larger masses in 
groups 2 and 3 also rotate more rapidly than field stars with similar 
masses and evolutionary states, but the differences decrease with 
increasing mass. The differences in mean rotation speed between h 
and $\chi$ Per and the field stars primarily reflect a paucity of slowly 
rotating stars in the double cluster, particularly among stars in groups 
1 and 2. It has been known for a long time (e.g. Slettebak 1968) that 
there are a large number of Be stars in h and $\chi$ Per, and so the fact 
that we find a bias toward rapidly rotating stars in these clusters is
perhaps not surprising.  We note as well that the observed distribution of 
rotation speeds in h and $\chi$ Per is unusual
compared with other, albeit lower density, clusters (e.g. Brown \& 
Verschueren (1997)). 

What causes the differences in N(\textit{v sin i}) between h and $\chi$ Per and 
the field?  Is the near absence of slow rotation seen in h and $\chi$ Per 
a consequence of a difference in the initial conditions that 
characterize the formation of stars in a dense, bound cluster as 
compared with the presumably lower density regions in which field 
stars form?  If so, what specific differences in initial conditions 
are the determining factors?  Why do the distributions of 
\textit{v sin i} for h 
and $\chi$ Per stars appear to converge progressively toward the 
distributions of \textit{v sin i} seen for the field stars in the 
two hotter regions?  Is this apparent convergence the result of 
processes that are 
effective after stars reach the ZAMS?  Or was the similarity of 
N(\textit{v sin i}) between h and $\chi$ Per and the field found for 
early B stars imposed at the time of star formation?

\subsection{Rotation Changes during Evolution away from the ZAMS}

We consider first the question of how evolution affects N(\textit{v sin i}) 
after stars reach the ZAMS, and specifically whether it is plausible 
that the massive stars in group 3 initially shared the high average 
rotation speeds of their cooler, lower mass cohorts in groups 1 and 2, 
but converged to the field star average as they evolved.

Heger \& Langer (2000) and Meynet \& Maeder (2000) have calculated 
models of evolving rotating stars for stars spanning the masses 
represented among group 3.  These models show that qualitatively, as 
high mass main sequence stars evolve from the ZAMS toward the end of core 
hydrogen burning, their surface rotation should decrease as a result 
of (a) changes in stellar moments of inertia; and (b) loss of stellar 
angular momentum via strong stellar winds.  However, the magnitude of 
the decrease in surface rotation rate is 
predicted to be modest because the loss of angular momentum from the 
surface layers is partially compensated by the transport of angular 
momentum from the core of the star.  For a 12 $M\sun$ star, Heger and Langer 
predict that the rotation rates of stars initially rotating at 
300 km s$^{-1}$ or less will decline by only 20-25\% during the 
course of their main sequence lifetimes; surface rotation begins to 
increase only after core hydrogen is exhausted and stars approach 
the terminal age main sequence (TAMS). Stars of this mass whose 
initial rotation 
speeds exceed 300 km/sec are predicted to slow by an additional 
10\% during the first $\sim$2 Myr after they reach the ZAMS.  

Similarly, for a 9 $M\sun$ star rotating initially at 300 km s$^{-1}$, Meynet 
and Maeder predict a 27\% decrease in \textit{v sin i} during main sequence 
evolution. Since the decrease of rotation speed as the star evolves from 
the ZAMS to a point just 
prior to the TAMS is predicted to be nearly identical in percentage terms for 
stars of differing initial masses and rotation rates (20-30\% over 
the range  mass range 9-12 $M\sun$) one might expect some convergence 
of \texttt{<}\textit{v sin i}\texttt{>} for two groups of stars, 
one of which initially contained 
a large number of rapid rotators and a second group of stars dominated by 
slow rotators.  However, the convergence  predicted from extant
models ($\sim$25\%) is much smaller than that required to reduce 
the mean \textit{v sin i} by a factor of 2, the amount required
to evolve a distribution, N(\textit{v sin i}), similar to that observed
for stars in group 1, to one closely resembling that found for group 3.

Observations as well argue against a factor of two decrease in rotation 
rates as stars evolve.  Wolff \& Preston (1978) and WEP searched 
for a correlation 
between \textit{v sin i} 
and age by sorting field B stars according to the Str\"{o}mgren  
$\beta$ index,  which through its sensitivity to surface gravity provides 
a measure of distance from the ZAMS and hence of age.  While this 
technique is somewhat uncertain since the $\beta$ index can be affected 
both by emission and by extremely rapid rotation, these authors found no 
significant systematic differences in 
\texttt{<}\textit{v sin i}\texttt{>} as a function 
of distance 
from the ZAMS, and hence argued that any change in \textit{v sin i} with age 
must be small at least for this heterogeneous sample. Abt (2003) 
looked for systematic differences in 
\texttt{<}\textit{v sin i}\texttt{>} between B dwarfs and giants 
of the same masses among  a sample of field stars. By using spectral type as 
a surrogate for mass, Abt reports that for stars of 9 $M\sun$ (the highest 
mass included in his study) rotation rates decline by only 11\% from 
class V to class III stars.  Perhaps the best extant evaluation 
of age-drive rotation changes  is that of Gies \& Huang (2003), who 
observed B0-B3 stars (the spectral type range populating our group 
3) in clusters with ages in the range 3-18 Myr.  They found evidence for a 
possible decrease in rotation of about 20\% from the ZAMS to ages 
t $\sim$10 Myr, followed by an apparent spinup of perhaps 30\% among stars 
older than 10 Myr; this spinup occurs well before the stars reach 
the TAMS and is not predicted by the models of single rotating stars.  

Both theoretical calculations and observations therefore argue against 
the hypothesis that the early B stars in h and $\chi$ Per initially 
rotated twice as rapidly as their counterparts in the field.  Rather, 
it appears that the differences between h and $\chi$ Per and the field 
stars are intrinsically largest among the late B stars and diminish 
with increasing mass.

Several authors over the years have reported results similar to those that 
we find here---namely that stars in clusters rotate more rapidly than 
stars in the field (e.g. Bernacca \& Perinotto 1974; WEP 1982; 
Gies \& Huang 2003; Keller 2004).  In the latter two cases, the difference 
was attributed to evolutionary effects. Gies \& Huang suggested that the 
field stars might represent a population that is somewhat older than 
their sample of fairly young cluster stars and that spin down 
processes reduce the average rotation rates of the field stars 
(as predicted by theory). By contrast,  Keller, who reports observations 
of rotation speeds among LMC clusters with ages greater than 10 Myr, 
suggests that the higher rotation speeds observed among cluster stars results 
from LMC clusters having ages systematically larger than their field 
counterparts. In this case, the spin up among the putatively older LMC 
cluster sample is attributed to the increase in 
\texttt{<}\textit{v sin i}\texttt{>} expected as 
stars approach the TAMS. 

Both studies selected their samples based primarily on spectral type; 
as a consequence, the age distributions among the field and cluster 
samples are not well defined.   In the current study, we have been 
careful to match the ages represented among our field and cluster stars, 
and in any case we find the largest differences between the field and 
h and $\chi$ Per samples among the late B stars, 
which are essentially unevolved.  Therefore, for the current sample, 
we cannot attribute the differences 
in N(\textit{v sin i}) to a systematic difference in age 
between the cluster stars 
in h and $\chi$ Per and the field stars.

\subsection{Initial Conditions and Rotation}

The above results suggest that the effects of evolution on observed 
\textit{v sin i} distributions should in principle be small. Hence, the observed 
differences between h and $\chi$ Per and the field seem most logically 
attributed to differences in initial conditions. Three different types 
of initial conditions have been cited as factors that influence the 
observed angular momentum of stars:  1) binary frequency; 2) composition; 
and 3) environment.

\subsubsection{Binary Characteristics}

Binary frequency has the potential to influence the rotation of the 
component stars in two ways.  First, closely spaced binaries are 
expected to have their rotational and orbital motions synchronized.  
Second, the formation of a binary system, whatever its spacing, may 
result in a system in which most of the angular momentum resides in 
orbital motions rather than stellar rotation.  

ALG examined the issue of synchronization in their sample of field B stars.  
They conclude that stars with orbital periods less than 2.4 days 
rotate synchronously and that stars with periods between 2.4-5 days 
are synchronized within a factor of two.  Moreover, they find that the 
average rotation observed among stars in close binaries is indeed lower 
than for apparently single stars, a result expected if orbital and 
rotational speeds are synchronized (note that for a period of 2.4 days, 
the corresponding rotational velocity of a typical B star would be 
60 km s$^{-1}$). However, ALG warn that the apparent difference in 
rotation speeds between binary and single stars might also reflect the 
fact that stars with published orbits are biased toward sharp-lined 
stars.  For their sample of cluster stars, Gies \& Huang also find 
a difference in the same sense:  
\texttt{<}\textit{v sin i} \texttt{>} = 125 km s$^{-1}$ for stars known to 
have variable velocities 
and 168 km s$^{-1}$ for constant stars.

In order to attribute the more rapid rotation seen among the h 
and $\chi$ Per sample to a difference in the effect of synchronization, 
h and $\chi$ Per would have to be deficient in close binaries.  The 
data presented in Table 1 provide an estimate of the number of stars 
in our h and $\chi$ Per sample with instantaneous observed amplitudes
K \texttt{>} 30 km s$^{-1}$ based on a comparison of (typically) a single observed
radial velocity with the cluster mean. We can use the complete 
orbital information for the field binaries to estimate what fraction 
of the field stars would be detected as binaries in a single observation 
by calculating for each star for which an orbit is known the fraction 
of time that the observed velocity differs from the average by more 
than 30 km s$^{-1}$.  Since 
stars with amplitudes greater than 30 km s$^{-1}$ are all fairly close 
binaries, we have made the simplifying assumption that their orbits 
are circular.  Column 4 of Table 5 shows the fraction of the field stars 
either with known orbits and velocity amplitudes 
K \texttt{>} 30 km/sec or that ALG 
reported to have double lines.  This fraction refers to the number of
binary systems; that is, we have counted each binary pair as one 
system.  Column 4 of this table also shows the fraction of binaries 
with known orbits that would have been detected with a single observation
to have a velocity that differed by more than 30 km s$^{-1}$ from 
the center of mass velocity.
This number is to be compared with the fraction in h and $\chi$ Per.  
Given the small number statistics for the h and $\chi$ Per sample, 
there is no evidence for a significant difference in the number 
of short period binaries between h and $\chi$ Per  and the field. 

As an additional check on our conclusion that a deficiency of close 
binaries in h and $\chi$ Per is not the explanation for their higher 
mean rotation speeds, we can ask how 
\texttt{<}\textit{v sin i}\texttt{>} for the field stars 
would change if the field sample contained no close binaries.  Reference 
to Table 5 shows that the fraction of large amplitude 
(K \texttt{>} 30 km s$^{-1}$), 
and therefore close, binaries among the field stars
in our three mass intervals ranges 
from 10-18\%.  If we calculate \texttt{<}\textit{v sin i}\texttt{>} 
for the total sample including 
binaries and the sample excluding those stars with velocity variations 
greater than 30 km s$^{-1}$, the two averages for each of the three 
mass intervals differ by no more than 6\% (see Table 4).  Based on our 
analysis of the field star sample, we therefore conclude that 
even if h and $\chi$ Per contained no close binaries, the difference in 
average rotation speeds between cluster and field samples cannot 
be explained.

Although the difference in rotation speeds between h and $\chi$ Per and 
field stars cannot be explained  by a difference in the fraction of 
synchronized (close) binaries,  recent work by Brown and 
Verschueren (1997) suggests that the overall frequency of binaries 
could influence the distribution of \textit{v sin i}. These 
authors conclude that 
the binary stars in the loose Sco OB2 association rotate on average 
more slowly than single stars. However, most of the detected binaries 
in this survey are so widely separated that one would not expect 
synchronization to be effective.  Brown \& Verschueren suggest instead 
that the observed slow rotation among the Sco OB2 binaries is a result of 
preferential allocation of angular momentum to orbital motion rather 
than stellar rotation during the star formation process. It could be 
that in a dense region like h and $\chi$ Per, the formation of widely 
separated binaries is somehow inhibited by dynamical interactions and as a 
result there is a deficiency of widely separated binaries relative to what is 
seen in the field.  While such an effect could explain the more 
rapid rotation seen in h and $\chi$ Per, there is at present no way of 
testing this hypothesis. 

\subsubsection{Chemical Composition}

Studies of rotation among B stars in the Magellanic Clouds have led 
to the suggestion that stars formed in the lower metalllicity LMC and 
SMC rotate more rapidly than stars formed in the solar neighborhood.  
Indirect evidence cited in support of this hypothesis is 
the observed anti-correlation between the frequency of Be stars, which 
are rapid rotators, and metallicity (Maeder, Grebel, \& Mermilliod 1999). 
More recently, Keller (2004) has obtained rotational velocities for 
(a) B0-B2 stars in LMC clusters that have ages between 10 and 30 Myr; 
and (b) LMC field stars in the same range of spectral types. He concluded 
that both the LMC field and cluster samples rotate more rapidly than 
their field star counterparts in the Milky Way; the difference is 
significant at the 2 $\sigma$ level. 

Differing abundances cannot, however, account for the fact that h 
and $\chi$ Per stars on average rotate more rapidly than Galactic field 
stars.  Vrancken et al. (2000) find that the abundances measured for 
early B giants in h and $\chi$ Per are in reasonable agreement with abundances 
measured by other authors for main sequence B stars, including nearby 
field stars.

\subsubsection{Possible Relationship between Initial Conditions and 
ZAMS Rotation Speeds}

Field B stars are generally assumed to have formed in loose clusters, 
associations, or aggregates that disperse rapidly. Since the h 
and $\chi$ Per clusters are still bound, it is reasonable to assume 
that the stars comprising these clusters were formed in regions of 
higher average protostellar densities than those characterizing the 
birthplace of a typical field B star. For reasons noted in 
the introduction, theory suggests that time-averaged protostellar 
accretion rates are likely to be higher in denser regions.  If the 
linkage between initial density and protostellar accretion rate 
predicted by theory is correct, current models predict higher rotation 
speeds among the outcome stellar populations formed in dense regions. 

Suppose that (a) material from a protostellar core infalls onto a disk; 
(b) material is transported to a forming star through a circumstellar 
accretion disk; and (c) the inner region of the accretion disk is 
linked to the star via stellar magnetospheric field lines as described 
in the introduction. If so, various formulations of the accretion 
process (e.g. Johns-Krull \& Gafford 2002) predict that, for fixed 
magnetic field strength, the rotation rate should vary directly 
as a (positive) power of the accretion rate, since higher 
accretion rates tend to 
crush the stellar magnetosphere and drive the disk/magnetosphere 
corotation radius of the disk closer to the surface of the star.  The 
rotation rate at the birthline, i.e. when the phase of rapid accretion 
ends, also depends directly on a power of the protostellar radius, which 
in turn also depends on the accretion rate, with larger accretion rates 
leading to larger radii along the birthline. The combination of higher 
initial angular velocity and higher initial radius for higher 
accretion rates, can in principle lead to higher ZAMS rotation 
speeds following pre-main 
sequence contraction from the birthline to the ZAMS.  This notional 
linkage---high protostellar accretion rates which lead to high 
angular rotation 
speeds along the birthline and, finally, high ZAMS rotation 
speeds---neglects many important effects:  probable complex topology 
of the magnetic field, the localized structures of the accretion columns, 
and the differential rotation of the disk and the star, which leads to 
disconnection of the magnetic field that links the two and reduces the 
spin-down torque (Matt \& Pudritz 2004). Nevertheless, by making 
plausible assumptions regarding  stellar accretion rates 
in low density star-forming regions, Wolff, Strom, \& Hillenbrand (2004) 
demonstrate that it is possible to account for the observed trends 
in specific angular momentum with mass for stars in the mass range 
0.1-10 $M\sun$.

The large difference in rotation speeds between groups 1 and 2 in h 
and $\chi$ Per and their field analogs may find explanation in the 
sensitivity of the mass-radius relationship along the stellar birthline 
to accretion rate over the mass range $M \sim 4-12 M\sun$., i.e. the mass 
range observed among the h and $\chi$ Per and field B star sample 
discussed above.  Over this range of stellar masses, models of the 
accretion process predict: (a) that the radius of the forming star 
when it is deposited on the birthline is larger for larger 
accretion rates; but (b) perhaps most significantly, that for all 
accretion rates, the mass-radius relationship exhibits a sharp 
maximum  in radius at a mass somewhere in the range 4-12 $M\sun$. The 
local maximum in radius reflects the onset of deuterium shell burning, 
which produces a substantial expansion in the radius of the accreting 
protostar; the specific range of masses at which D-shell burning sets in, 
and the radius expands, depends in turn on the protostellar accretion rate. 
The potential significance of a local maximum in radius along the 
birthline for stars having masses somewhere between 4-12 $M\sun$ results 
from the fact that the rotation rates observed when B stars reach the 
ZAMS depend on the initial rotation rates at the birthline and the 
spinup as stars contract from the birthline to the ZAMS. 

We speculate that the large difference between h and $\chi$ Per stars 
and the field among groups 1 and 2 reflects  a difference in the mass 
accretion rates characterizing the cluster (high accretion rate) and 
field birthplaces (low accretion rate) that in turn produces differences 
in initial radii along the birthline that are largest among stars 
having masses in the range 4.5 to 9 $M\sun$. If correct, stars in this
mass range should show the largest difference in rotation speeds---a 
direct result of greater spinup during contraction toward the ZAMS. At 
higher masses, the difference in rotation speeds between cluster and 
field should be smaller, since the initial radii along the birthline 
are similar or, for those stars where the birthline meets the ZAMS, identical.

\section{Conclusions}

Observations of rotational velocities show that B stars in the h and 
$\chi$ Per double cluster rotate on average more rapidly than field stars 
of the same mass and age. This result combined with other similar 
results in the literature clearly establishes that the rotation rates 
of B stars \textit{differ significantly} among stars born in different regions. 
We have argued that the observed differences are 
\textit{likely} built in at 
the time the stars were formed and not a consequence of subsequent 
evolutionary processes. Since h and $\chi$ Per are currently bound 
clusters, it seems reasonable to assume that stars in these clusters 
were formed in much denser environments than field B stars of 
comparable age, which have presumably escaped from the loose 
clusters, associations, or small groups in which they formed. We 
have identified two possible explanations for the observed 
differences in rotation speeds, each related to the density at the 
time of star formation. One hypothesis is essentially untestable 
with current facilities: that h and $\chi$ Per are deficient in wide 
binaries relative to the field and that when wide binaries are 
formed, much of the available angular momentum appears as orbital 
rather than stellar rotational angular momentum. The second 
possibility is that protostellar accretion is more rapid in higher 
density regions, and that high accretion rates lead to more rapid 
rotation---a consequence of the hypothesis that ``disk locking""
accounts for initial stellar angular momenta. This hypothesis is 
attractive because the effects of differences in accretion rate 
during the stellar assembly phase are predicted to be particularly 
dramatic among B stars. For stars in the mass range 4-12 $M\sun$ , the 
initial radius along the stellar birthline reflects the effects of 
deuterium shell burning, which causes a substantial expansion of the 
star; the amount of the expansion and the range of stellar masses 
over which it occurs both increase with increasing accretion rates. 
The combination of high initial accretion rate with high initial 
radius followed by contraction from the birthline to the ZAMS can 
lead to high ZAMS rotation speeds compared to stars of similar mass 
formed in regions characterized by lower accretion rates.

We emphasize that our results for h and $\chi$ Per, while suggestive,
do not provide conclusive evidence of a direct relationship between
environment and outcome rotational velocities. With
the advent of 8- to 10-m telescopes, it should be possible to test 
the hypothesis that outcome stellar rotation speeds are in fact linked to 
initial stellar density among stars born in a wide variety of 
environments by observing a large sample of stars located in young 
clusters both the in Milky Way and in the Magellanic Clouds. Such 
observations will provide the basis for establishing robustly both 
the cosmic dispersion in N(vsini) as well as systematic differences 
attributable to initial stellar density, chemical composition, or 
other parameters. For regions of sufficient youth, it may also be 
possible to search for differences in the location of the birthline. 
If the predicted correlations between rotation speed, birthline 
location, and environment can be found, it would be possible for the 
first time to link the initial conditions under which star formation 
takes place and to outcome observables---a crucial first step toward 
a predictive theory of star formation.


Facilities: \facility{WIYN}.




\clearpage



\begin{figure}
\plotone{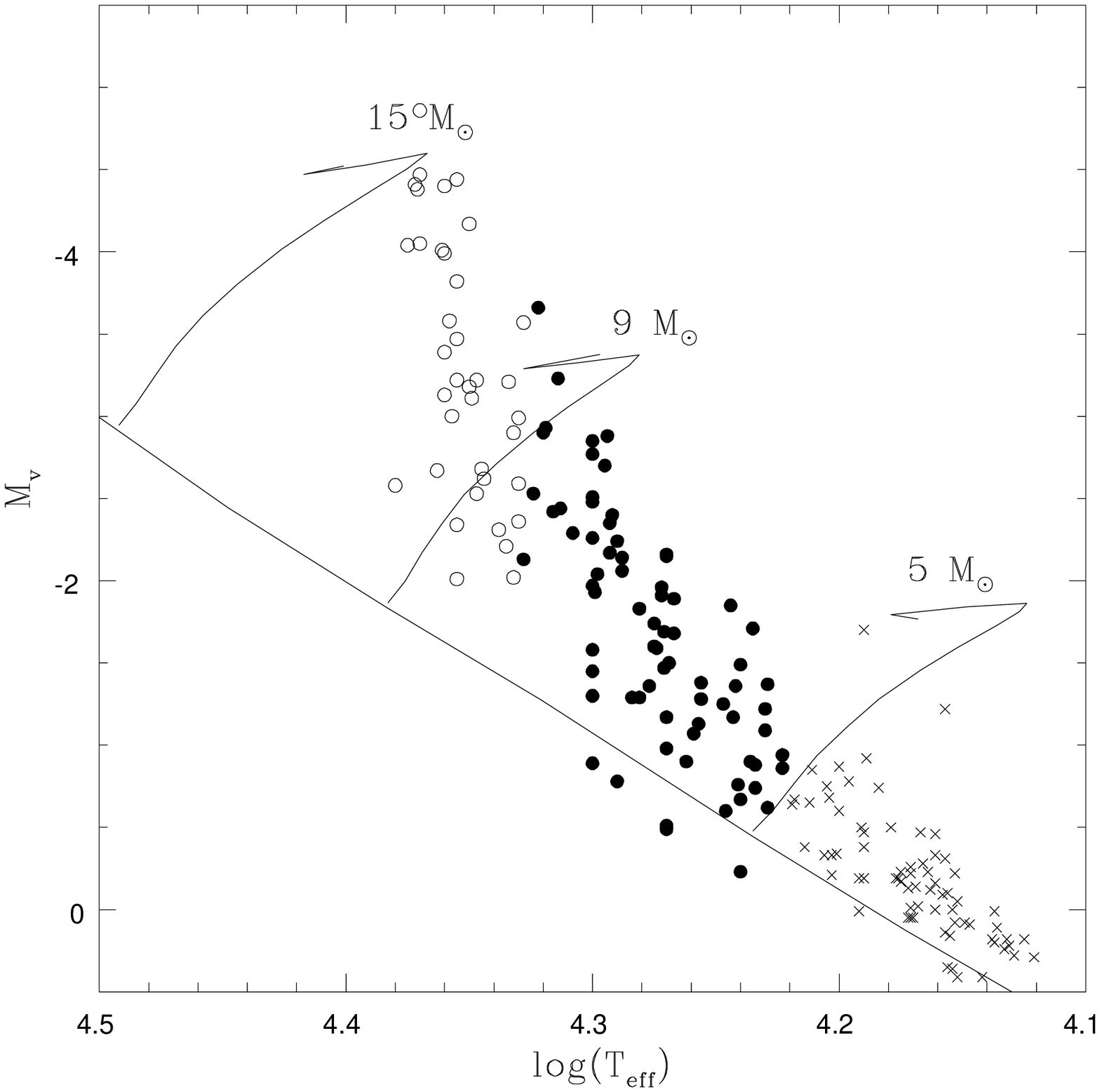}
\caption{The M$_{v}$ vs. log T$_{eff}$ diagram for the stars in h and
$\chi$ Per.  The data are from Slesnick, Hillenbrand, \& Massey (2001).  
For analysis, these stars have been divided according to T$_{eff}$ 
into three groups, represented from hottest to coolest by the open 
circles, filled circles, and crosses, respectively.  Also shown are the
ZAMS and evolutionary tracks for models representing stars with 
masses of 15, 9, and 5 $M\sun$ from Schaller et al. (1992).}
\end{figure}

\clearpage
\begin{figure}
\plottwo{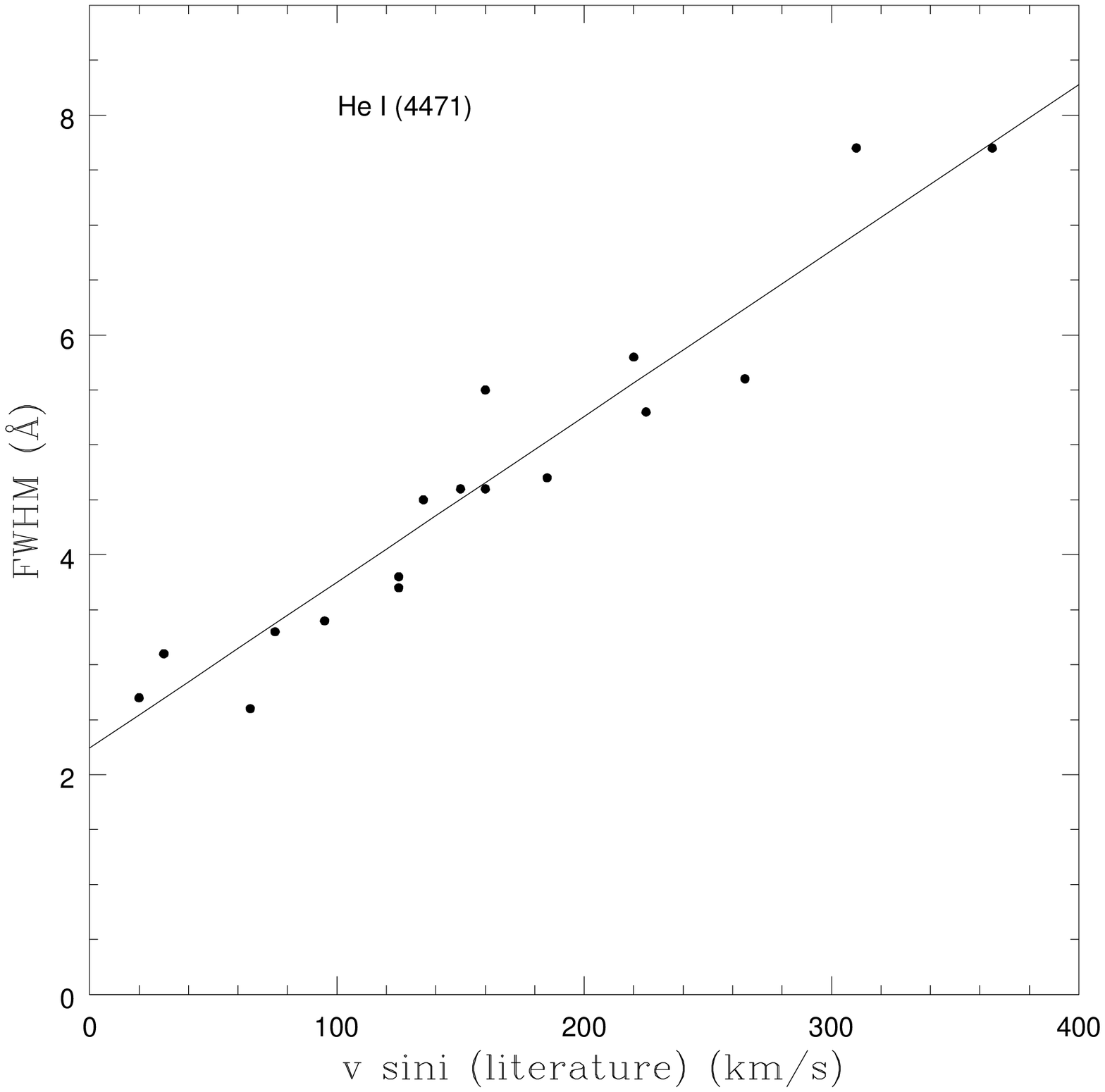}{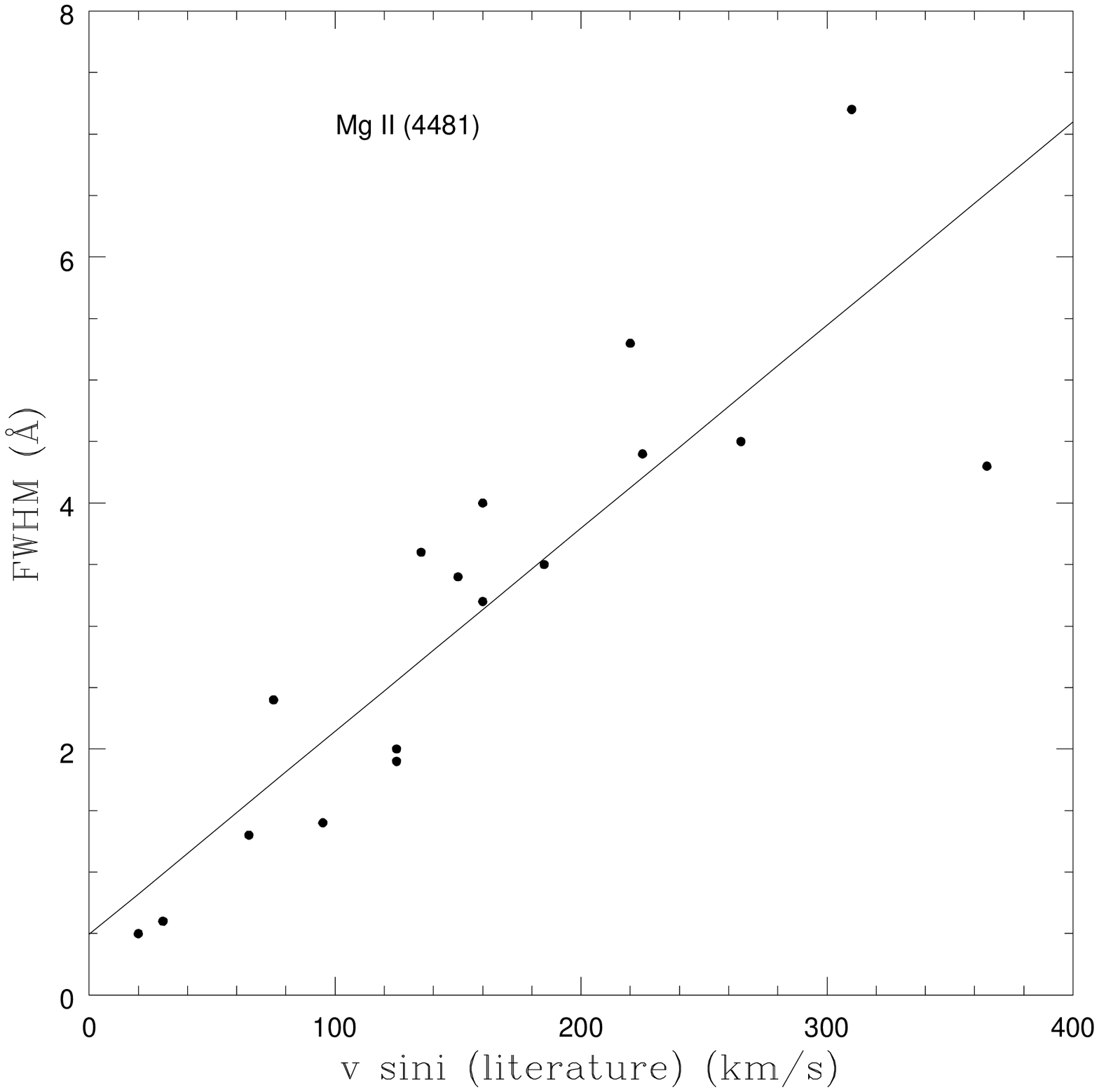}
\caption{The relationships between FWHM of He I $\lambda$ 4471 (left)
and Mg II $\lambda$ 4481 (right) and \textit{v sin i} for standard
stars in I Lac (Abt \& Hunter 1962).  The best least squares fit
to the data is shown in each panel.}
\end{figure}

\clearpage
\begin{figure}
\plotone{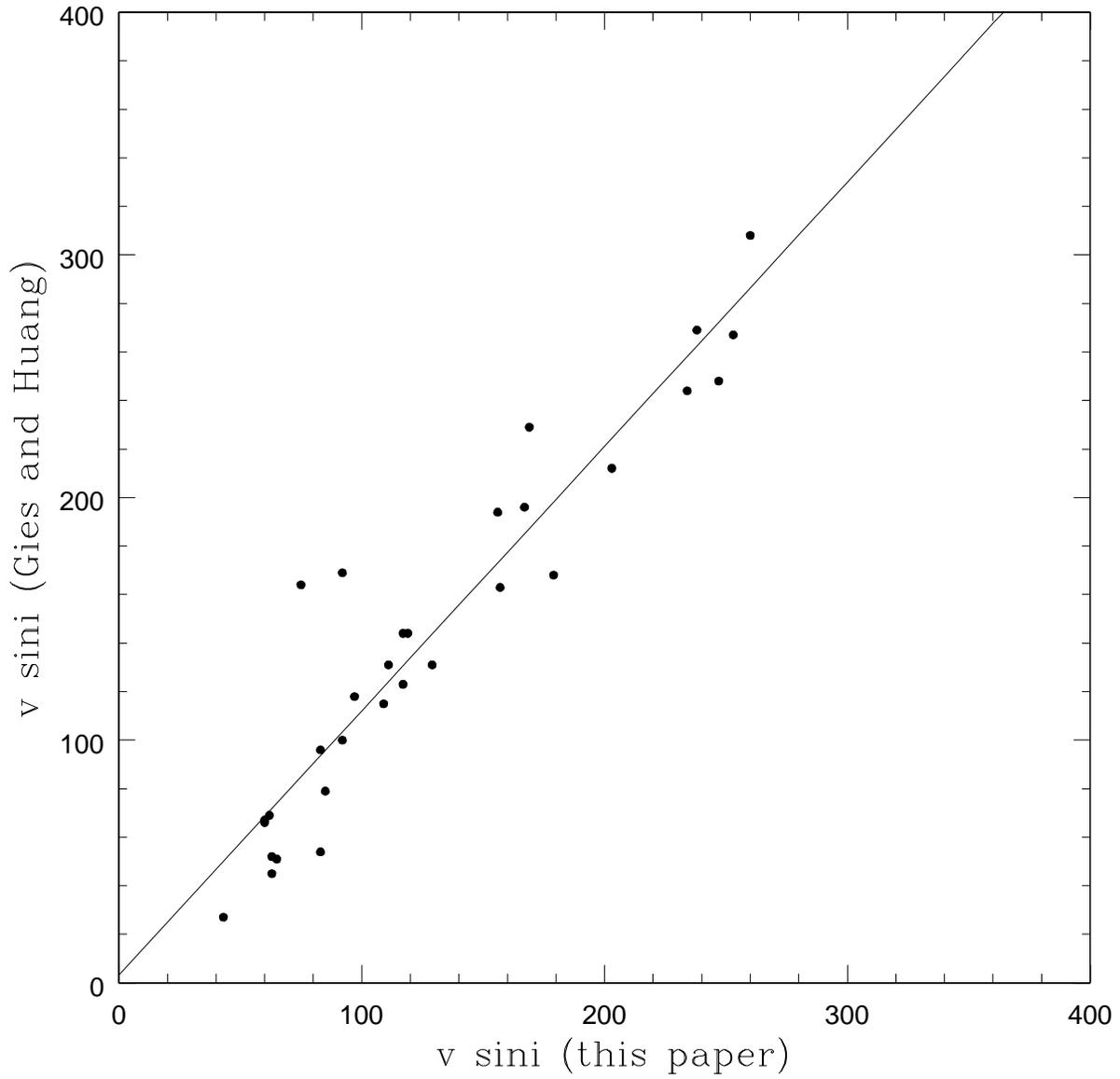}
\caption{The comparison between the values of \textit{v sin i} measured
in the current study and those derived by Gies \& Huang (in preparation)
for stars in h and $\chi$ Per that are common to the two programs.
The best fitting straight line is given by $v sin i (Gies \& Huang)
=1.05 v sin i (current study) + 11 km s^{-1}$.}
\end{figure}

\clearpage
\begin{figure}
\plotone{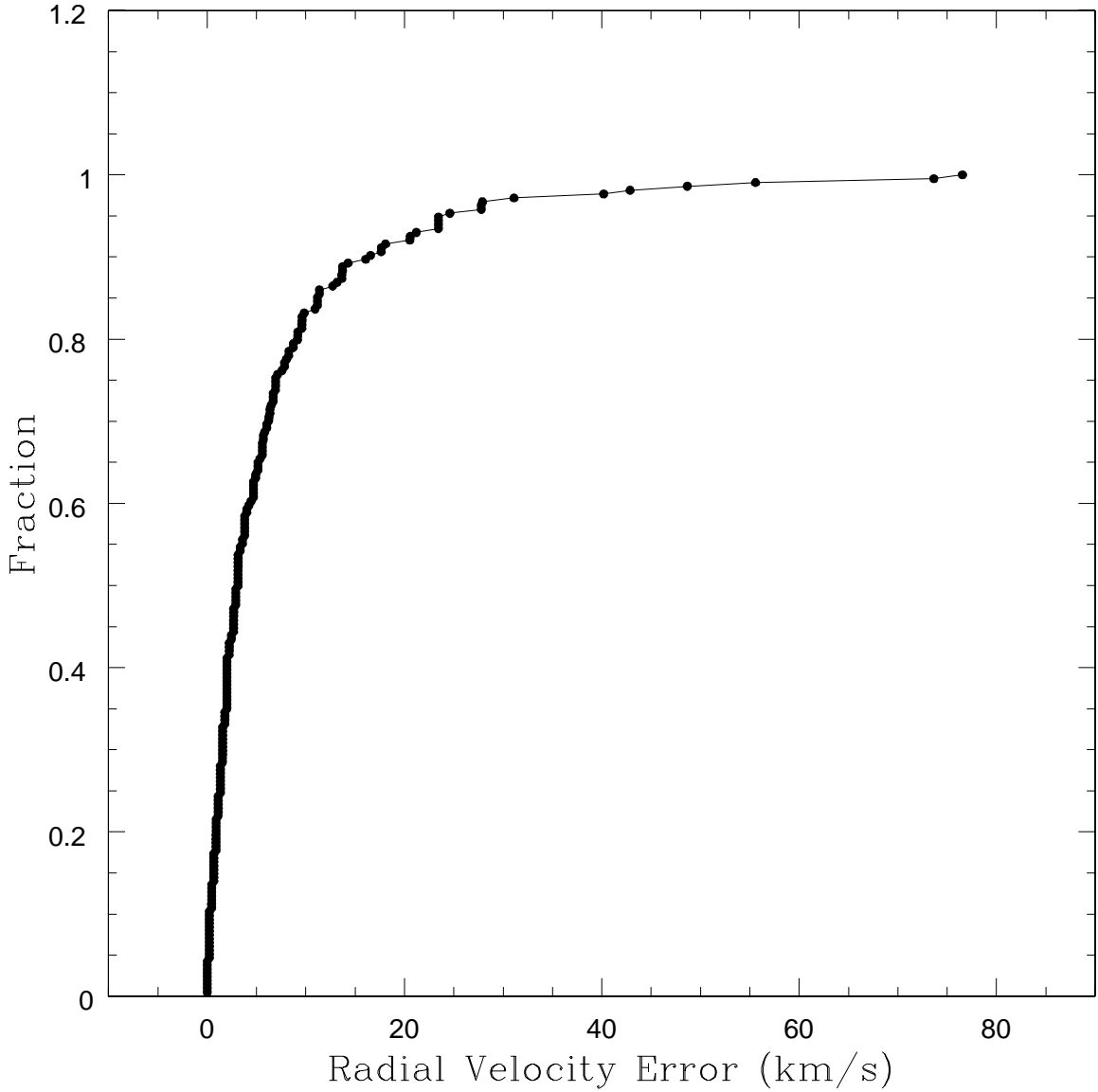}
\caption{The cumulative distribution of velocity differences between pairs
of exposures of the same stars taken less than 45 minutes apare.  Since
radial velocity should not change significantly over this short
interval of time, the differences can be taken as an estimate
of the measurement error.  Note that only about 3\% of the stars have velocity
differences greater than 30 km s$^{-1}$.}
\end{figure}

\clearpage
\begin{figure}
\plotone{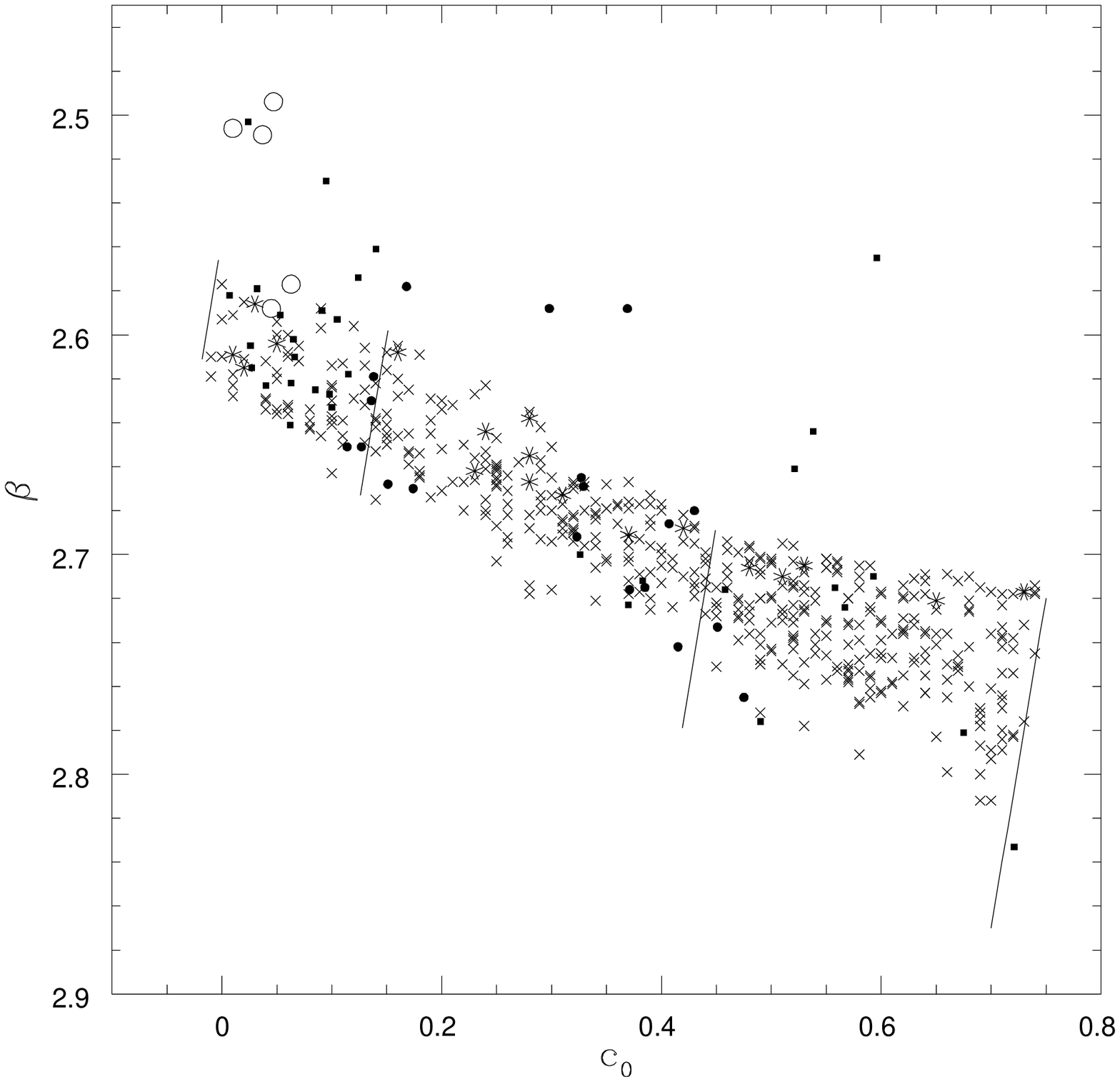}
\caption{Measured values of $\beta$ and c$_{0}$ for ALG field stars 
included in our sample (crosses). The 
boundaries between groups are shown as solid lines. Dots
represent measurements for that subset of the
h and $\chi$ Per sample with published photometry;
open circles indicate known emission line objects in h and $\chi$. The outliers for the h and $\chi$ Per
sample may represent either objects having unreported hydrogen line emission, objects rotating at 0.9 breakup
speed but viewed at a modest inclination angle (Collins \& Sonneborn, 1977),
or lower quality measurements.}
\end{figure}

\clearpage
\begin{figure}
\plotone{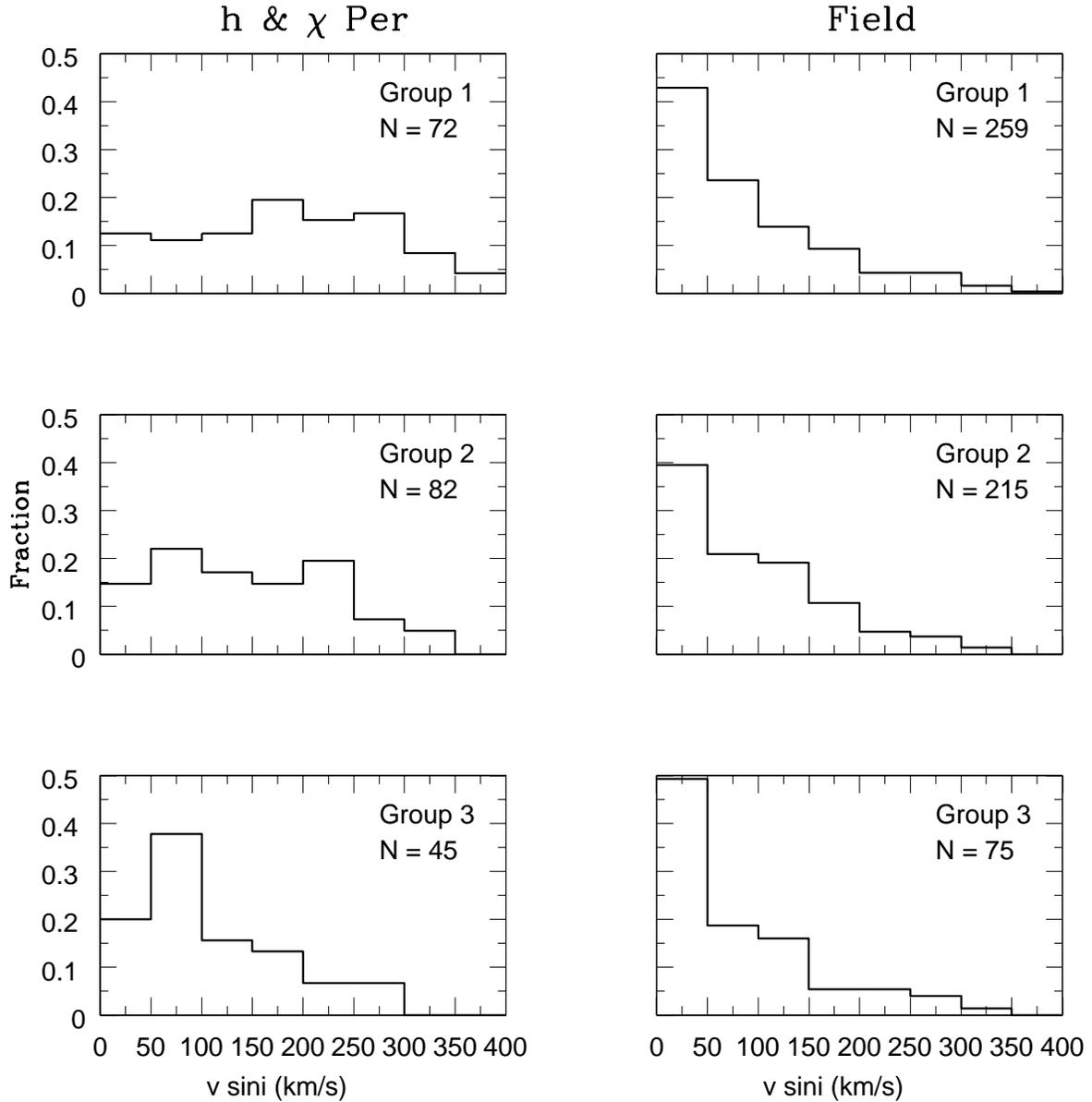}
\caption{Distribution of \textit{v sin i} for each of the three
groups of stars in h and $\chi$ Per compared with the distributions for
field stars.  Note the large differences in the distributions for the
stars in the coolest group (Group 1) and the progressive convergence
of the distributions with increasing temperature.}
\end{figure}

\clearpage
\begin{figure}
\plotone{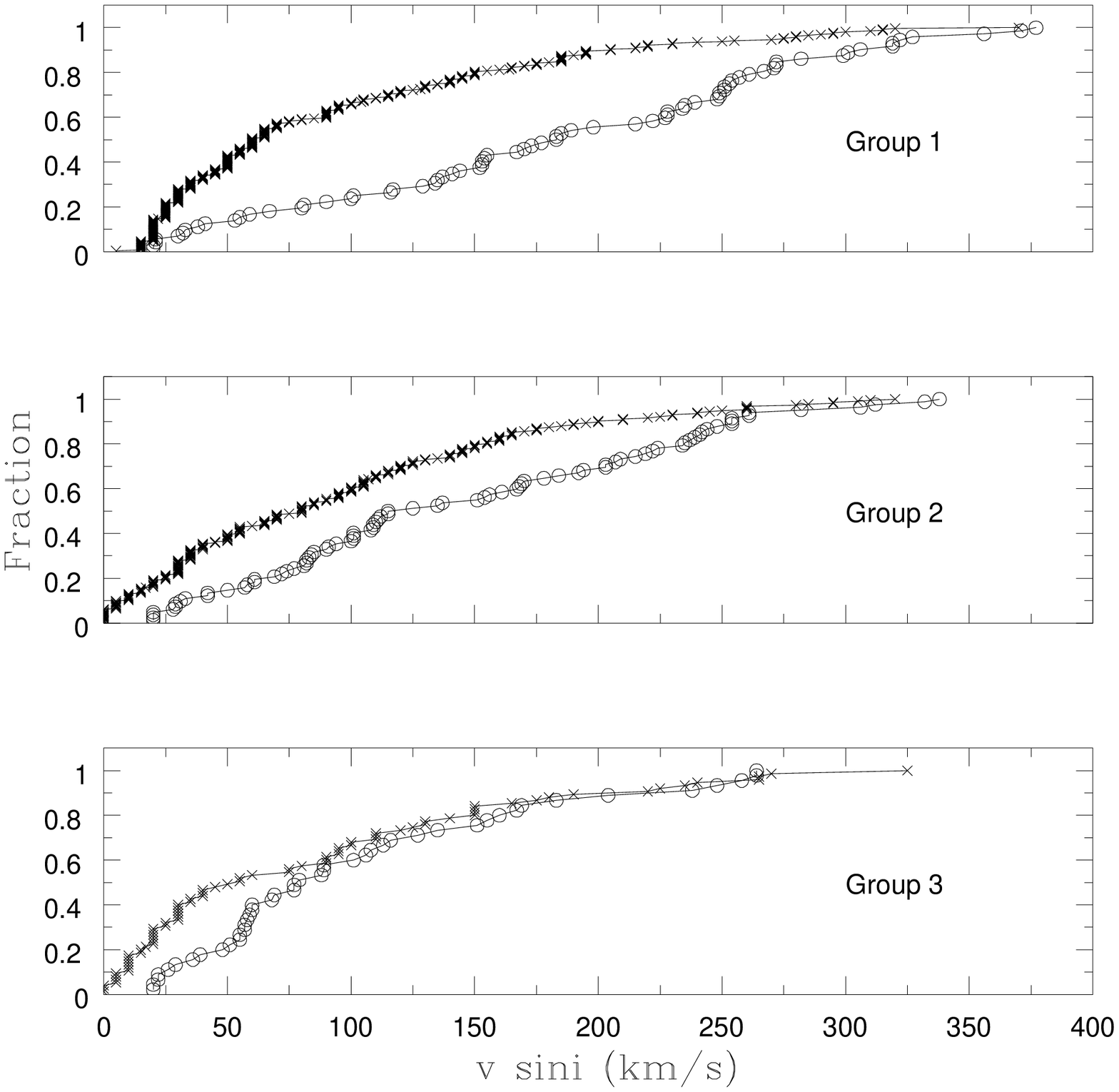}
\caption{Cumulative distribution of \textit{v sin i} for the stars 
in h and $\chi$ Per (open circles) compared with the distributions for 
field stars (crosses) of similar temperature and age. Stars of Group 1
comprise late B stars, little evolved from the ZAMS. Stars of Group 2
comprise middle B stars, evolved from the ZAMS by less than 1.0 mag.
Stars of Group 3 comprise early B stars, significantly evolved from
the ZAMS. In all cases, the fraction of stars with \textit{v sin i} \texttt{<} 
100 km s$^{-1}$ is larger
in h and $\chi$ Per, with the largest difference occurring for
Group 1, i.e. for late B stars.}
\end{figure}

\clearpage








\clearpage


\begin{deluxetable}{rrrrlrrcrrc}
\rotate
\tabletypesize{\scriptsize}




\tablecaption{Data for Stars in h and $\chi$ Persei}

\tablenum{1}

\tablehead{\colhead{ID} & \colhead{Oosterhoff} & \colhead{M$_{v}$} & \colhead{Log(T$_{eff}$)} & \colhead{Sp. Type} & \colhead{Mass} & \colhead{\textit{v sin i}} & \colhead{Group} & \colhead{c$_{0}$} & \colhead{$\beta$} & 
\colhead{Binarity} \\ 
\colhead{} & \colhead{} & \colhead{} & \colhead{} & \colhead{} & 
\colhead{(Sun$=$1)} & \colhead{(km s$^{-1}$)} & \colhead{} & \colhead{} & \colhead{} & \colhead{} } 

\startdata
9 & 2227 & -5.57 & 4.375 & B1.5II & 21 & 22 & --- & 0.031 & 2.584 &   \\
12 & 662 & -4.86 & 4.37 & B1.5I & 16.4 & 22 & 3 & 0.007 & 2.582 &  \\
16 & 1132 & -5.12 & 4.34 & B2I & 16.8 & 60 & --- &  &  &  \\
19 & 2296 & -4.79 & 4.34 & B2I & 15.1 & 146 & --- & 0.05 & 2.598 &  \\
28 & 847 & -4.47 & 4.37 & B1.5Ie & 15.2 & 55 & 3 & 0.063 & 2.577 &  \\
32 & 2299 & -4.05 & 4.37 & B1.5I & 13.2 & 106 & 3 & 0.027 & 2.615 &  \\
33 & 2371 & -4.44 & 4.355 & B2III & 14 & 60 & 3 & 0.024 & 2.503 & delta v = -62 \\
36 & 2541 & -4.4 & 4.36 & B1.5III & 14.4 & 55 & 3 & 0.066 & 2.61 &  \\
40 & 1116 & -4.65 & 4.45 & B0.5V & 20 & 93 & --- & 0.088 & 2.635 &  \\
41 & 2088 & -4.41 & 4.372 & B1IIIe & 14.8 & 89 & 3 & 0.01 & 2.506 &  \\
42 & 717 & -4.09 & 4.405 & B1V & 15 & 57 & 3 & 0.026 & 2.605 & delta v = 40 \\
43 & 843 & -4.17 & 4.35 & B1.5V & 13.2 & 88 & 3 & 0.091 & 2.589 &  \\
46 & 1268 & -4.38 & 4.371 & B0.5V & 14.6 & 127 & 3 & 0.032 & 2.579 &  \\
47 & 2311 & -4.04 & 4.375 & B1.5II & 13.3 & 20 & 3 & 0.065 & 2.602 &  \\
49 & 692 & -3.81 & 4.44 & B0.5I & 15.5 & 169 & 3 &  &  \\
52 & 782 & -3.99 & 4.36 & B1.5III & 12.7 & 36 & 3 & 0.095 & 2.53 \\
54 &  & -4.01 & 4.361 &  & 12.9 & 116 & 3 &  &  \\
57 & 922 & -4.15 & 4.397 & B2V & 14.2 & 258 & 3 & 0.053 & 2.591 \\
58 & 1261 & -4.72 & 4.502 & B3Ve & 24 & 261 & --- &  &  \\
66 & 2284 & -4.24 & 4.484 & Be & 19.7 & 299 & --- & -0.082 & 2.381 \\
71 & 980 & -3.82 & 4.355 & B2III & 12 & 39 & 3 & 0.098 & 2.627 \\
72 & 1078 & -3.81 & 4.405 & B1V & 14.1 & 167 & 3 & 0.1 & 2.633 \\
77 &  & -3.56 & 4.418 &  & 13.8 & 108 & 3 &  &  \\
78 & 1364 & -3.66 & 4.322 &  & 11.1 & 203 & 2 &  &  \\
81 & 864 & -3.57 & 4.328 & B2V & 10.8 & 113 & 3 & 0.14 & 2.561 \\
83 & 2246 & -3.47 & 4.355 & B2III & 11.5 & 101 & 3 & 0.04 & 2.623 \\
87 & 517 & -3.25 & 4.267 & Be & 8.8 & 178 &  &  &  \\
91 &  & -3.58 & 4.358 &  & 11.3 & 264 & 3 &  &  \\
96 & 2165 & -3.5 & 4.405 & B1Ve & 13.2 & 79 & 3 & 0.047 & 2.494 \\
100 & 1161 & -3.73 & 4.405 & B1Ve & 13.8 & 135 & 3 & 0.045 & 2.588 \\
102 &  & -3.11 & 4.349 &  & 10.6 & 58 & 3 &  &  &  \\
106 & 929 & -3.39 & 4.36 & B1.5III & 11.5 & 151 & 3 & 0.115 & 2.618 &  \\
109 &  & -3.23 & 4.314 &  & 9.6 & 161 & 2 &  &  &  \\
117 & 936 & -3.18 & 4.35 & B1.5V & 10.8 & 26 & 3 & 0.063 & 2.622 &  \\
120 & 1085 & -3.22 & 4.355 & B2III & 10.9 & 59 & 3 & 0.062 & 2.641 &  \\
121 &  & -3.22 & 4.347 &  & 10.8 & 20 & 3 &  &  &  \\
131 & 2262 & -3.13 & 4.36 & B1.5III & 10.9 & 204 & 3 & 0.124 & 2.574 &  \\
133 &  & -3.05 & 4.389 &  & 11.6 & 248 & 3 &  &  &  \\
138 &  & -2.9 & 4.32 &  & 9.5 & 115 & 2 &  &  &  \\
141 & 2566 & -3.21 & 4.334 & Be & 10.4 & 183 & 3 &  &  & delta v = -44 \\
144 & 978 & -2.99 & 4.33 & B2IV & 9.8 & 29 & 3 & 0.085 & 2.625 &  \\
149 &  & -2.9 & 4.332 &  & 9.8 & 77 & 3 &  &  &  \\
150 &  & -2.67 & 4.363 &  & 10.1 & 22 & 3 &  &  &  \\
158 & 859 & -2.85 & 4.3 & B2V & 9 & 192 & 2 & 0.168 & 2.578 &  \\
160 &  & -2.88 & 4.294 &  & 9 & 156 & 2 &  &  &  \\
163 &  & -2.93 & 4.319 &  & 9.5 & 20 & 2 &  &  &  \\
165 &  & -2.68 & 4.345 &  & 9.7 & 60 & 3 &  &  \\
171 & 622 & -2.34 & 4.355 & B2III & 9.4 & 57 & 3 &  &  \\
177 &  & -2.53 & 4.347 &  & 9.5 & 77 & 3 &  &  \\
178 & 2185 & -2.59 & 4.33 & B2IV & 9.2 & 155 & 3 & 0.326 & 2.7 \\
184 &  & -3 & 4.357 &  & 10.6 & 48 & 3 &  &  \\
188 &  & -2.42 & 4.316 &  & 8.6 & 33 & 2 &  &  \\
189 & 1282 & -2.58 & 4.38 & Be & 10.4 & 264 & 3 & 0.037 & 2.509 \\
191 & 1109 & -2.77 & 4.3 & B2V & 8.9 & 170 & 2 & 0.174 & 2.67 \\
193 & 919 & -2.97 & 4.294 &  & 8.5 & 94 & 2 &  &  \\
194 &  & -2.08 & 4.218 & B5II & 6.7 & 67 & --- &  &  \\
198 &  & -2.7 & 4.295 &  & 8.7 & 57 & 2 &  &  \\
200 & 963 & -2.26 & 4.3 & B2.5II & 8.1 & 29 & 2 & 0.114 & 2.651 \\
201 &  & -2.53 & 4.324 &  & 9 & 72 & 2 &  &  \\
202 & 2114 & -2.48 & 4.3 & B2V & 8.4 & 169 & 2 &  &  \\
203 & 2232 & -2.15 & 4.27 & B3V & 7.5 & 101 & 2 & 0.127 & 2.651 \\
205 &  & -2.35 & 4.293 &  & 8.1 & 215 & 2 &  &  \\
206 &  & -2.17 & 4.293 &  & 7.9 & 209 & 2 &  &  \\
208 & 1041 & -2.51 & 4.3 & B2V & 8.5 & 154 & 2 & 0.151 & 2.668 \\
219 & 892 & -2.24 & 4.29 & B2.5III & 7.9 & 61 & 2 & 0.136 & 2.63 \\
222 &  & -2.44 & 4.313 &  & 8.6 & 31 & 2 &  &  \\
223 &  & -2.21 & 4.335 &  & 8.8 & 238 & 3 &  &  \\
231 &  & -2.62 & 4.344 &  & 9.5 & 51 & 3 &  &  \\
235 & 950 & -2.36 & 4.33 & B2IV & 8.8 & 89 & 3 & 0.105 & 2.593 \\
237 &  & -2.4 & 4.292 &  & 8.2 & 312 & 2 &  &  \\
239 & 2229 & -2.16 & 4.27 & B3V & 7.5 & 235 & 2 &  &  \\
240 &  & -2.29 & 4.308 &  & 8.3 & 242 & 2 &  &  \\
241 &  & -2.13 & 4.328 &  & 8.5 & 137 & 2 &  &  \\
265 &  & -1.83 & 4.281 &  & 7.3 & 261 & 2 &  &  \\
277 &  & -2.14 & 4.288 &  & 7.8 & 109 & 2 &  &  \\
280 &  & -2.02 & 4.332 &  & 8.5 & 69 & 3 &  &  \\
284 &  & -2.01 & 4.355 & B2III & 9 & 68 & 3 &  &  \\
288 & 879 & -1.97 & 4.3 & B2V & 7.8 & 91 & 2 & 0.138 & 2.619 \\
289 &  & -1.96 & 4.272 &  & 7.3 & 194 & 2 &  &  &  \\
297 &  & -2.31 & 4.338 &  & 9 & 160 & 3 &  &  &  \\
298 &  & -1.93 & 4.299 &  & 7.7 & 112 & 2 &  &  & delta v = -41 \\
299 &  & -2.04 & 4.298 &  & 7.8 & 254 & 2 &  &  &  \\
314 & 2091 & -2.06 & 4.288 & Be & 7.7 & 167 & 2 &  &  &  \\
315 &  & -1.89 & 4.267 &  & 7.1 & 239 & 2 &  &  &  \\
318 &  & -1.85 & 4.244 &  & 6.8 & 110 & 2 &  &  &  \\
331 &  & -1.71 & 4.235 &   & 6.5 & 42 & 2 &  &  & SB2; delta v=106 \\
331 &  & -1.71 & 4.235 &   & 6.5 & 20 & 2 &  &  & SB2 \\
338 &  & -1.69 & 4.271 &  & 7 & 50 & 2 &  &  &  \\
341 &  & -1.74 & 4.275 &  & 7.1 & 178 & 2 &  &  &  \\
350 &  & -1.68 & 4.267 &  & 6.9 & 82 & 2 &  &  &  \\
359 & 2140 & -1.7 & 4.19 & B5V & 6 & 72 & --- &  &  &  \\
362 &  & -1.91 & 4.272 &  & 7.2 & 101 & 2 &  &  &  \\
363 &  & -1.47 & 4.271 &  & 6.8 & 338 & 2 &  &  &  \\
369 & 2301 & -1.58 & 4.3 & B2V & 7.4 & 135 & 2 &  &  & delta v = 31 \\
393 & 778 & -1.55 & 4.35 & B1.5V & 8.4 & 225 & & 0.293 & 2.571 &  \\
394 & 869 & -1.45 & 4.3 & B2V & 7.3 & 254 & 2 & 0.407 & 2.686 &  \\
401 & 952 & -1.36 & 4.242 &  & 6.3 & 219 & 2 & 0.323 & 2.692 &  \\
404 &  & -1.59 & 4.274 &  & 7 & 42 & 2 &  &  &  \\
409 &  & -1.5 & 4.269 &  & 6.8 & 237 & 2 &  &  &  \\
415 &  & -1.49 & 4.24 &  & 6.4 & 125 & 2 &  &  &  \\
431 &  & -1.29 & 4.284 &  & 6.8 & 207 & 2 &  &  &  \\
436 & 2379 & -1.3 & 4.3 & B2V & 7.1 & 244 & 2 &  &  &  \\
439 & 800 & -1.3 & 4.3 & B2V & 7.2 & 234 & 2 & 0.298 & 2.588 &  \\
443 &  & -1.6 & 4.275 &   & 7 & 28 & 2 &  &  & SB2; delta v=111 \\
443 &  & -1.6 & 4.275 &   & 7 & 20 & 2 &  &  & SB2 \\
444 &  & -0.78 & 4.196 &  & 5.3 & 267 & 1 &  &  &  \\
450 &  & -1.25 & 4.247 &  & 6.3 & 77 & 2 &  &  &  \\
452 & 2352 & -1.22 & 4.23 & B3III & 6 & 74 & 2 &  &  &  \\
453 &  & -1.36 & 4.277 &  & 6.8 & 109 & 2 &  &  &  \\
460 &  & -1.29 & 4.281 &  & 6.8 & 69 & 2 &  &  &  \\
494 &  & -0.96 & 4.129 &  & 4.9 & 169 & --- &  &  &  \\
497 &  & -1.09 & 4.23 &  & 5.9 & 101 & 2 &  &  &  \\
500 &  & -1.07 & 4.259 &  & 6.3 & 241 & 2 &  &  &  \\
501 &  & -1.38 & 4.256 &  & 6.5 & 100 & 2 &  &  &  \\
506 &  & -0.9 & 4.236 &  & 5.9 & 90 & 2 &  &  & delta v = -67 \\
507 &  & -1.28 & 4.256 &   & 6.4 & 29 & 2 &  &  & SB2; delta v=104 \\
507 &  & -1.28 & 4.256 &   & 6.4 & 20 & 2 &  &  & SB2 \\
513 & 896 & -1.17 & 4.35 & B1.5V & 7.9 & 403 & --- & 0.344 & 2.649 &  \\
514 &  & -1.22 & 4.157 &  & 5.2 & 21 & --- &  &  &  \\
517 &  & -1.37 & 4.229 &  & 6.1 & 111 & 2 &  &  &  \\
520 & 956 & -1.17 & 4.243 &   & 6.2 & 83 & 2 & 0.371 & 2.716 &  \\
531 &  & -0.74 & 4.234 &  & 5.7 & 168 & 2 &  &  &  \\
533 &  & -0.98 & 4.323 &  & 7.2 & 196 & --- &  &  &  \\
537 &  & -0.92 & 4.189 &  & 5.3 & 137 & 1 &  &  &  \\
540 & 965 & -1.17 & 4.27 & B3V & 6.5 & 82 & 2 & 0.329 & 2.669 &  \\
543 &  & -0.74 & 4.184 &   & 5.1 & 20 & 1 &  &  & SB2; delta v=126 \\
543 &  & -0.74 & 4.184 &   & 5.1 & 20 & 1 &  &  & SB2 \\
549 &  & -1.13 & 4.257 &  & 6.3 & 248 & 2 &  &  &  \\
551 &  & -0.87 & 4.2 &  & 5.4 & 271 & 1 &  &  &  \\
554 &  & -0.75 & 4.205 &  & 5.4 & 141 & 1 &  &  &  \\
562 &  & -0.86 & 4.286 &  & 6.5 & 81 & 2 &  &  &  \\
565 &  & -0.88 & 4.234 &  & 5.8 & 61 & 2 &  &  &  \\
576 & 876 & -0.98 & 4.27 & B3V & 6.4 & 115 & 2 & 0.327 & 2.665 &  \\
587 &  & -0.94 & 4.223 &  & 5.7 & 306 & 2 &  &  &  \\
590 &  & -0.5 & 4.179 &  & 4.9 & 53 & 1 &  &  &  \\
594 &  & -0.62 & 4.229 &  & 5.6 & 85 & 2 &  &  &  \\
598 &  & -0.46 & 4.161 &  & 4.7 & 215 & 1 &  &  &  \\
600 &  & -0.65 & 4.212 &  & 5.4 & 32 & 1 &  &  &  \\
602 & 2349 & -0.78 & 4.29 & B2.5V & 6.5 & 261 & 2 & 0.43 & 2.68 &  \\
611 &  & -0.5 & 4.116 &  & 4.4 & 237 & --- &  &  &  \\
616 &  & -0.9 & 4.262 &  & 6.2 & 254 & 2 &  &  &  \\
628 &  & -0.86 & 4.223 &  & 5.7 & 58 & 2 &  &  &  \\
632 &  & -0.85 & 4.211 &  & 5.5 & 248 & 1 &  &  &  \\
641 &  & -0.64 & 4.219 &  & 5.5 & 299 & 1 &  &  &  \\
650 & 872 & -0.89 & 4.3 & B2V & 6.8 & 332 & 2 & 0.369 & 2.588 &  \\
655 & 2297 & -0.47 & 4.19 & B5V & 5 & 254 & 1 & 0.383 & 2.712 &  \\
658 & 2211 & -0.49 & 4.27 & B3V & 6 & 282 & 2 & 0.475 & 2.765 &   \\
670 &  & -0.22 & 4.153 &  & 4.5 & 30 & 1 &  &  &  \\
675 & 659 & -0.38 & 4.19 & B5V & 5 & 153 & 1 &  &  &  \\
681 &  & -0.47 & 4.167 &  & 4.8 & 239 & 1 &  &  &  \\
683 & 2345 & -0.51 & 4.27 & B3V & 6 & 108 & 2 & 0.415 & 2.742 &  \\
687 &  & -0.76 & 4.241 &  & 5.8 & 203 & 2 &  &  &  \\
701 &  & -0.33 & 4.206 &  & 5.1 & 67 & 1 &  &  &  \\
709 & 923 & -0.21 & 4.203 &  & 5 & 38 & 1 & 0.37 & 2.723 &  \\
721 & 2275 & -0.6 & 4.246 &  & 5.8 & 184 & 2 & 0.385 & 2.715 &  \\
726 &  & -0.33 & 4.161 &  & 4.7 & 33 & 1 &  &  &  \\
736 & 1066 & -0.6 & 4.2 &  & 5.2 & 153 & 1 & 0.458 & 2.716 &  \\
749 &  & -0.38 & 4.214 &  & 5.2 & 251 & 1 &  &  &  \\
753 &  & -0.21 & 4.123 &  & 4.3 & 207 & --- &  &  \\
754 & 820 & -0.23 & 4.164 & BV & 4.6 & 319 & 1 & 0.521 & 2.661 \\
765 &  & -0.68 & 4.204 &  & 5.3 & 185 & 1 &  &  \\
767 &  & -0.31 & 4.157 &  & 4.6 & 177 & 1 &  &  \\
768 &  & -0.67 & 4.218 &  & 5.5 & 144 & 1 &  &  \\
774 &  & -0.67 & 4.24 &  & 5.7 & 151 & 2 &  &  \\
778 &  & -0.58 & 4.274 &  & 6.1 & 84 & 2 &  &  \\
782 & 2267 & -0.19 & 4.19 & B5V & 4.8 & 154 & 1 & 0.49 & 2.776 \\
783 &  & 0.01 & 4.137 &  & 4.3 & 117 & 1 &  &  \\
792 & 2350 & -0.23 & 4.24 &  & 5.4 & 222 & 2 & 0.451 & 2.733 \\
814 &  & -0.07 & 4.114 &  & 4.1 & 219 & --- &  &  \\
853 &  & -0.34 & 4.201 &  & 5.1 & 101 & 1 &  &  \\
854 &  & -0.26 & 4.171 &  & 4.7 & 155 & 1 &  &  \\
859 &  & -0.5 & 4.191 &  & 5 & 198 & 1 &  &  \\
862 &  & -0.17 & 4.175 &  & 4.7 & 90 & 1 &  &  \\
870 &  & -0.09 & 4.158 &  & 4.5 & 377 & 1 &  &  \\
878 &  & -0.22 & 4.171 &  & 4.7 & 282 & 1 &  &  & delta v = -43 \\
884 &  & -0.13 & 4.172 &  & 4.6 & 235 & 1 &  &  &  \\
892 &  & -0.16 & 4.161 &  & 4.6 & 183 & 1 &  &  &  \\
893 &  & -0.28 & 4.166 &  & 4.7 & 253 & 1 &  &  & delta v = -38 \\
896 & 804 & -0.1 & 4.156 &  & 4.5 & 301 & 1 & 0.538 & 2.644 &  \\
903 &  & -0.19 & 4.192 &  & 4.9 & 59 & 1 &  &  &  \\
918 & 2111 & -0.14 & 4.169 &   & 4.6 & 173 & 1 & 0.567 & 2.724 &  \\
920 &  & -0.19 & 4.177 &  & 4.7 & 81 & 1 &  &  &  \\
939 &  & 0 & 4.154 &  & 4.4 & 306 & 1 &  &  &  \\
940 &  & 0 & 4.161 &  & 4.5 & 167 & 1 &  &  &  \\
941 &  & 0.14 & 4.157 &  & 4.4 & 134 & 1 &  &  &  \\
942 &  & -0.33 & 4.203 &  & 5.1 & 222 & 1 &  &  &  \\
952 & 2116 & -0.12 & 4.163 & B8V & 4.6 & 116 & 1 & 0.593 & 2.71 &  \\
954 & 830 & -0.25 & 4.27 & B3V & 5.8 & 224 & 2 &  &  &  \\
970 & 784 & -0.02 & 4.168 & B8V & 4.6 & 228 & 1 & 0.596 & 2.565 &  \\
973 &  & -0.19 & 4.176 &  & 4.7 & 170 & 1 &  &  &  \\
986 &  & 0.09 & 4.147 &  & 4.3 & 322 & 1 &  &  &  \\
1007 &  & -0.23 & 4.175 &  & 4.7 & 319 & 1 &  &  & delta v = -41: \\
1009 &  & 0.05 & 4.171 &  & 4.5 & 327 & 1 &  &  &  \\
1017 &  & 0.18 & 4.125 &  & 4.1 & 189 & 1 &  &  &  \\
1018 &  & -0.05 & 4.152 &  & 4.4 & 80 & 1 &  &  & delta v = -37: \\
1040 &  & 0.08 & 4.153 &  & 4.4 & 272 & 1 &  &  &  \\
1053 &  & -0.01 & 4.171 &  & 4.6 & 356 & 1 &  &  & delta v = 76 \\
1102 &  & 0.18 & 4.132 &  & 4.2 & 257 & 1 &  &  &  \\
1124 &  & 0.22 & 4.131 &  & 4.1 & 249 & 1 &  &  &  \\
1128 & 2426 & 0.28 & 4.129 & B8V & 4.1 & 261 & 1 &  &  & delta v = 32 \\
1140 &  & 0.08 & 4.149 &  & 4.3 & 249 & 1 &  &  &  \\
1141 &  & -0.11 & 4.223 &  & 5.1 & 272 & 1 &  &  &  \\
1169 &  & 0.35 & 4.156 &  & 4.2 & 55 & 1 &  &  &  \\
1170 &  & 0.2 & 4.137 &  & 4.2 & 251 & 1 &  &  &  \\
1172 & 1077 & 0.11 & 4.136 &  & 4.2 & 129 & 1 & 0.675 & 2.781 &  \\
1176 &  & 0.01 & 4.192 &  & 4.8 & 41 & 1 &  &  &  \\
1216 &  & 0.24 & 4.133 &  & 4.1 & 228 & 1 &  &  & delta v = -44: \\
1218 &  & 0.05 & 4.17 &  & 4.5 & 371 & 1 &  &  &  \\
1234 & 1049 & 0.05 & 4.172 &   & 4.5 & 152 & 1 & 0.558 & 2.715 &  \\
1250 &  & 0.18 & 4.138 &  & 4.2 & 234 & 1 &  &  &  \\
1271 &  & 0.29 & 4.121 &  & 4 & 100 & 1 &  &  &  \\
1280 &  & 0.16 & 4.155 &  & 4.3 & 21 & 1 &  &  &  \\
1294 &  & 0.36 & 4.154 &  & 4.2 & 135 & 1 &  &  &  \\
1403 &  & 0.41 & 4.142 &  & 4.1 & 227 & 1 &  &  &  \\
1405 & 1118 & 0.32 & 4.128 &  & 4 & 183 & 1 & 0.721 & 2.833 &  \\
1518 &  & 0.41 & 4.152 &  & 4.2 & 21 & 1 &  &  & delta v = 50 \\
\enddata




\end{deluxetable}


\begin{deluxetable}{rrlrrrc}
\tabletypesize{\scriptsize}




\tablecaption{Data for Field Stars}

\tablenum{2}

\tablehead{\colhead{HR} & \colhead{HD} & \colhead{Sp. Type} & \colhead{\textit{v sin i}} & \colhead{$\beta$} & \colhead{c$_{0}$} & \colhead{} \\ 
\colhead{} & \colhead{} & \colhead{} & \colhead{(km s$^{-1}$)} & \colhead{} & \colhead{} & \colhead{} } 

\startdata
Group 1  &  &  &  &  &  &  \\
15 & 358 & B8IVmnp & 50 & 2.743 & 0.52 &  \\
28 & 584 & B7IV & 15 & 2.757 & 0.55 \\
61 & 1256 & B6III/IV & 150 & 2.741 & 0.49 &  \\
70 & 1438 & B8V & 20 & 2.751 & 0.67 &  \\
78 & 1606 & B7V & 104 & 2.737 & 0.52 &  \\
96 & 2054 & B9IV & 40 & 2.775 & 0.69 &  \\
121 & 2729 & B6V & 95 & 2.709 & 0.49 &  \\
123 & 2772 & B8Vn & 220 & 2.754 & 0.71 &  \\
137 & 3038 & B9III & 195 & 2.778 & 0.69 &  \\
144 & 3240 & B7III & 60 & 2.75 & 0.67 &  \\
149 & 3322 & B8IIImnp & 25 & 2.732 & 0.58 &  \\
223 & 4636 & B9III & 95 & 2.739 & 0.58 &  \\
326 & 6676 & B8V & 120 & 2.742 & 0.71 &  \\
345 & 6972 & B9IV & 100 & 2.812 & 0.70 &  \\
348 & 7019 & B7III-IV & 60 & 2.741 & 0.65 &  \\
364 & 7374 & B8III & 25 & 2.747 & 0.60 &  \\
369 & 7546 & B9IIIsp & 25 & 2.719 & 0.62 &  \\
438 & 9298 & B7IIImnp & 50 & 2.739 & 0.60 &  \\
491 & 10425 & B8IIIn & 240 & 2.715 & 0.69 &  \\
561 & 11857 & B5III & 25 & 2.717 & 0.54 &  \\
562 & 11905 & B8III & 40 & 2.71 & 0.58 &  \\
612 & 12767 & B9.5sp & 45 & 2.714 & 0.51 &  \\
677 & 14272 & B8V & 25 & 2.763 & 0.64 &  \\
682 & 14392 & B9sp & 90 & 2.772 & 0.49 &  \\
702 & 14951 & B7IV & 215 & 2.727 & 0.51 &  \\
746 & 16004 & B9mnp & 30 & 2.753 & 0.56 &  \\
760 & 16219 & B5V & 30 & 2.72 & 0.49 &  \\
785 & 16727 & B7IIIp & 20 & 2.729 & 0.47 &  \\
811 & 17081 & B7IV & 25 & 2.717 & 0.60 &  \\
836 & 17543 & B6V & 70 & 2.703 & 0.48 &  \\
846 & 17743 & B8III & 50 & 2.741 & 0.57 &  \\
847 & 17769 & B7V & 145 & 2.741 & 0.54 &  \\
873 & 18296 & B9p & 25 & 2.767 & 0.58 &  \\
890 & 18537 & B7V & 90 & 2.739 & 0.52 &  \\
896 & 18604 & B6III & 105 & 2.72 & 0.57 &  \\
910 & 18883 & B7V & 65 & 2.758 & 0.57 &  \\
954 & 19832 & B6IV-V & 110 & 2.746 & 0.55 &  \\
982 & 20315 & B8V & 185 & 2.736 & 0.65 &  \\
1038 & 21364 & B9Vn & 195 & 2.783 & 0.65 &  \\
1047 & 21455 & B7V & 120 & 2.731 & 0.55 &  \\
1051 & 21551 & B8V & 295 & 2.746 & 0.67 &  \\
1094 & 22316 & B9p & 150 & 2.778 & 0.53 &  \\
1097 & 22402 & B8Vn & 320 & 2.731 & 0.60 &  \\
1100 & 22470 & B8/B9III & 128 & 2.728 & 0.47 &  \\
1113 & 22780 & B7Vne & 230 & 2.705 & 0.53 &  \\
1140 & 23288 & B7IV & 185 & 2.749 & 0.63 &  \\
1141 & 23300 & B6V & 20 & 2.702 & 0.49 &  \\
1144 & 23324 & B8V & 185 & 2.748 & 0.64 &  \\
1145 & 23338 & B6IV & 105 & 2.702 & 0.55 &  \\
1146 & 23363 & B7V & 140 & 2.735 & 0.64 &  \\
1172 & 23753 & B8V & 290 & 2.737 & 0.71 &  \\
1202 & 24388 & B8V & 135 & 2.736 & 0.61 &  \\
1207 & 24504 & B6V & 130 & 2.721 & 0.47 &  \\
1213 & 24587 & B5V & 30 & 2.744 & 0.50 &  \\
1243 & 25330 & B5V & 140 & 2.724 & 0.45 &  \\
1305 & 26670 & B5Vn & 270 & 2.715 & 0.46 &  \\
1307 & 26676 & B8Vn & 175 & 2.748 & 0.58 &  \\
1315 & 26793 & B9Vn & 275 & 2.732 & 0.73 &  \\
1328 & 27026 & B9V & 220 & 2.783 & 0.72 &  \\
1363 & 27563 & B5III & 35 & 2.703 & 0.52 &  \\
1375 & 27742 & B8IV-V & 175 & 2.763 & 0.64 &  \\
1377 & 27777 & B8V & 250 & 2.744 & 0.64 &  \\
1397 & 28114 & B6IV & 25 & 2.698 & 0.46 &  \\
1399 & 28149 & B7V & 115 & 2.726 & 0.48 &  \\
1402 & 28217 & B8IV & 60 & 2.717 & 0.53 &  \\
1424 & 28503 & B8V & 60 & 2.73 & 0.55 &  \\
1445 & 28929 & B9p & 55 & 2.749 & 0.53 &  \\
1449 & 29009 & B9sp & 55 & 2.701 & 0.49 &  \\
1469 & 29335 & B7V & 65 & 2.75 & 0.51 &  \\
1484 & 29589 & B8IV & 70 & 2.728 & 0.45 &  \\
1576 & 31373 & B9V & 70 & 2.731 & 0.52 &  \\
1600 & 31764 & B7V & 140 & 2.71 & 0.68 &  \\
1610 & 32040 & B9Vn & 295 & 2.77 & 0.69 &  \\
1671 & 33224 & B8V & 155 & 2.755 & 0.64 &  \\
1696 & 33802 & B8V & 185 & 2.755 & 0.59 &  \\
1705 & 33949 & B7V & 120 & 2.717 & 0.70 &  \\
1750 & 34762 & B9IV & 230 & 2.743 & 0.72 &  \\
1757 & 34863 & B7/B8V & 285 & 2.734 & 0.62 &  \\
1759 & 34880 & B8III & 45 & 2.724 & 0.59 &  \\
1769 & 35104 & B8II & 95 & 2.725 & 0.47 &  \\
1791 & 35497 & B7III & 60 & 2.703 & 0.56 &  \\
1860 & 36589 & B6V & 90 & 2.728 & 0.52 &  \\
1902 & 37098 & B9IV-V & 50 & 2.768 & 0.58 &  \\
1920 & 37320 & B8III & 25 & 2.736 & 0.64 &  \\
1945 & 37646 & B8IV & 130 & 2.762 & 0.60 &  \\
1957 & 37808 & B9.5IIIsp & 30 & 2.722 & 0.45 &  \\
1985 & 38478 & B8IIImnp & 55 & 2.705 & 0.58 &  \\
1997 & 38670 & B9Vn & 215 & 2.745 & 0.59 &  \\
2038 & 39417 & B9V & 145 & 2.75 & 0.66 &  \\
2109 & 40574 & B8IIIn & 205 & 2.709 & 0.64 &  \\
2116 & 40724 & B8V & 130 & 2.776 & 0.73 &  \\
2127 & 40964 & B8V & 115 & 2.782 & 0.72 &  \\
2130 & 41040 & B8III & 140 & 2.718 & 0.60 &  \\
2139 & 41269 & B9sp & 85 & 2.766 & 0.71 &  \\
2202 & 42657 & B9mnp & 65 & 2.737 & 0.55 &  \\
2207 & 42784 & B8Vnn & 370 & 2.72 & 0.57 &  \\
2223 & 43153 & B7V & 65 & 2.756 & 0.57 &  \\
2237 & 43362 & B9III & 145 & 2.77 & 0.71 &  \\
2248 & 43526 & B7III & 75 & 2.696 & 0.52 &  \\
2297 & 44766 & B8IIIn & 170 & 2.745 & 0.60 &  \\
2374 & 46075 & B6III & 50 & 2.736 & 0.62 &  \\
2438 & 47395 & B7III & 65 & 2.714 & 0.53 &  \\
2454 & 47756 & B8IIIs & 30 & 2.715 & 0.58 &  \\
2461 & 47964 & B8III & 45 & 2.718 & 0.72 &  \\
2497 & 49028 & B8IV & 45 & 2.72 & 0.49 &  \\
2519 & 49606 & B7III & 20 & 2.702 & 0.50 &  \\
2521 & 49643 & B8IIIn & 255 & 2.723 & 0.52 &  \\
2522 & 49662 & B7IV & 90 & 2.751 & 0.45 &  \\
2589 & 51104 & B8Vn & 160 & 2.753 & 0.67 &  \\
2605 & 51688 & B8III & 35 & 2.705 & 0.53 &  \\
2613 & 51892 & B7III & 50 & 2.704 & 0.56 &  \\
2676 & 53929 & B9.5III & 25 & 2.708 & 0.53 &  \\
2760 & 56446 & B8III & 185 & 2.725 & 0.65 &  \\
2826 & 58346 & B8/B9V & 165 & 2.754 & 0.72 &  \\
2860 & 59136 & B5III & 65 & 2.712 & 0.64 &  \\
2922 & 60863 & B8V & 185 & 2.756 & 0.59 &  \\
2947 & 61554 & B6V & 280 & 2.712 & 0.52 &  \\
2949 & 61556 & B5IVn & 70 & 2.75 & 0.57 &  \\
2956 & 61672 & B7V & 280 & 2.726 & 0.59 &  \\
3201 & 68099 & B6III & 50 & 2.72 & 0.62 &  \\
3353 & 71997 & B4V & 15 & 2.72 & 0.47 &  \\
3470 & 74604 & B8V & 150 & 2.759 & 0.61 &  \\
3500 & 75333 & B9mnp & 35 & 2.747 & 0.63 &  \\
3607 & 77665 & B8V & 90 & 2.738 & 0.72 &  \\
3652 & 79158 & B8IIImnp & 60 & 2.706 & 0.55 &  \\
3982 & 87901 & B7V & 300 & 2.723 & 0.71 &  \\
4119 & 90994 & B6V & 80 & 2.73 & 0.48 &  \\
4493 & 101391 & B9p & 50 & 2.765 & 0.59 &  \\
4696 & 107348 & B8V & 195 & 2.738 & 0.71 &  \\
4857 & 111226 & B8V & 70 & 2.73 & 0.60 &  \\
4967 & 114376 & B7III & 115 & 2.712 & 0.52 &  \\
5250 & 121847 & B8V & 165 & 2.757 & 0.66 &  \\
5407 & 126769 & B8V & 195 & 2.753 & 0.57 &  \\
5475 & 129174 & B9p & 25 & 2.745 & 0.54 &  \\
5597 & 133029 & B9sp & 30 & 2.8 & 0.69 &  \\
5614 & 133529 & B7V & 310 & 2.732 & 0.57 &  \\
5655 & 134946 & B8III & 150 & 2.738 & 0.52 &  \\
5733 & 137391 & F0V & 90 & 2.745 & 0.74 &  \\
5780 & 138764 & B6IV & 20 & 2.72 & 0.50 &  \\
5910 & 142250 & B6Vp & 15 & 2.75 & 0.49 &  \\
6003 & 144844 & B9V & 20 & 2.791 & 0.58 &  \\
6023 & 145389 & B9mnp & 20 & 2.787 & 0.69 &  \\
6054 & 146001 & B8V & 90 & 2.748 & 0.49 &  \\
6079 & 146926 & B8V & 180 & 2.736 & 0.70 &  \\
6294 & 152909 & B7/B8III & 50 & 2.752 & 0.57 &  \\
6340 & 154204 & B7IV/V & 275 & 2.725 & 0.51 &  \\
6520 & 158704 & B9II/III & 20 & 2.757 & 0.57 &  \\
6545 & 159376 & Ap & 164 & 2.726 & 0.68 &  \\
6567 & 159975 & B8II-IIImnp & 95 & 2.718 & 0.74 &  \\
6720 & 164447 & B8Vne & 170 & 2.721 & 0.65 &  \\
6919 & 169990 & B8III/IV & 110 & 2.764 & 0.71 &  \\
6967 & 171247 & B8IIIsp & 55 & 2.717 & 0.73 &  \\
6968 & 171301 & B8IV & 40 & 2.772 & 0.69 &  \\
6990 & 171961 & B8III & 55 & 2.721 & 0.54 &  \\
6997 & 172044 & B8II-IIIp & 30 & 2.696 & 0.48 &  \\
7035 & 173117 & B8III & 30 & 2.705 & 0.58 &  \\
7039 & 173300 & B8III & 35 & 2.733 & 0.71 &  \\
7073 & 173936 & B6V & 90 & 2.739 & 0.47 &  \\
7113 & 174933 & B9II-IIIp & 20 & 2.753 & 0.58 &  \\
7115 & 174959 & B6IV & 50 & 2.707 & 0.49 &  \\
7147 & 175744 & B9sp & 50 & 2.701 & 0.49 &  \\
7239 & 177817 & B7V & 130 & 2.742 & 0.68 &  \\
7241 & 177863 & B8V & 60 & 2.731 & 0.57 &  \\
7245 & 178065 & B9III & 15 & 2.718 & 0.71 &  \\
7248 & 178125 & B8III & 60 & 2.755 & 0.59 &  \\
7283 & 179527 & B9sp & 35 & 2.712 & 0.67 &  \\
7285 & 179588 & B9IV & 35 & 2.812 & 0.69 &  \\
7339 & 181558 & B5III & 20 & 2.715 & 0.44 &  \\
7346 & 181828 & B9V & 145 & 2.747 & 0.61 &  \\
7358 & 182255 & B6III & 25 & 2.736 & 0.48 &  \\
7361 & 182308 & B9mnp & 15 & 2.699 & 0.47 &  \\
7395 & 183056 & B9sp & 35 & 2.714 & 0.52 &  \\
7401 & 183339 & B8IVwe & 45 & 2.706 & 0.48 &  \\
7437 & 184606 & B8IIIn & 185 & 2.709 & 0.64 &  \\
7447 & 184930 & B5III & 55 & 2.707 & 0.56 &  \\
7452 & 184961 & B9sp & 40 & 2.789 & 0.70 &  \\
7457 & 185037 & B8Vne & 315 & 2.717 & 0.73 &  \\
7466 & 185268 & B5V & 195 & 2.694 & 0.46 &  \\
7493 & 186122 & B9III & 20 & 2.729 & 0.63 &  \\
7543 & 187235 & B8Vn & 315 & 2.76 & 0.68 &  \\
7593 & 188293 & B7Vn & 190 & 2.73 & 0.51 &  \\
7607 & 188651 & B6V+ & 150 & 2.715 & 0.52 &  \\
7608 & 188665 & B5V & 105 & 2.715 & 0.45 &  \\
7642 & 189432 & B5IV & 15 & 2.723 & 0.48 &  \\
7664 & 190229 & B9mnp & 20 & 2.695 & 0.51 &  \\
7721 & 192276 & B7V & 25 & 2.745 & 0.54 &  \\
7737 & 192659 & B9IV-V & 20 & 2.769 & 0.62 &  \\
7786 & 193722 & B9sp & 35 & 2.711 & 0.63 &  \\
7814 & 194636 & B4V & 30 & 2.763 & 0.60 &  \\
7840 & 195483 & B8V & 140 & 2.759 & 0.53 &  \\
7852 & 195810 & B6III & 50 & 2.702 & 0.55 &  \\
7878 & 196426 & B8IIIp & 20 & 2.735 & 0.62 &  \\
7885 & 196606 & B8IIIn & 50 & 2.729 & 0.62 &  \\
7922 & 197226 & B6III & 90 & 2.736 & 0.54 &  \\
7961 & 198174 & B7IIIp & 65 & 2.705 & 0.59 &  \\
7963 & 198183 & B5Ve & 100 & 2.71 & 0.51 &  \\
7978 & 198513 & B8np & 175 & 2.761 & 0.70 &  \\
8022 & 199578 & B5V & 30 & 2.71 & 0.46 &  \\
8033 & 199728 & Ap & 55 & 2.732 & 0.63 &  \\
8065 & 200614 & B8III & 20 & 2.714 & 0.74 &  \\
8094 & 201433 & B9V & 25 & 2.789 & 0.71 &  \\
8106 & 201834 & B9III & 25 & 2.732 & 0.57 &  \\
8109 & 201888 & B7III & 5 & 2.725 & 0.53 &  \\
8118 & 202149 & B9mnp & 35 & 2.78 & 0.71 &  \\
8141 & 202753 & B5V & 20 & 2.703 & 0.48 &  \\
8158 & 203206 & B6IV & 60 & 2.755 & 0.52 &  \\
8161 & 203245 & B6V & 75 & 2.731 & 0.50 &  \\
8218 & 204428 & B6V & 125 & 2.726 & 0.53 &  \\
8226 & 204754 & B8III & 20 & 2.709 & 0.66 &  \\
8240 & 205087 & B9sp & 30 & 2.799 & 0.66 &  \\
8292 & 206540 & B5IV & 20 & 2.719 & 0.50 &  \\
8338 & 207516 & B8V & 95 & 2.784 & 0.71 &  \\
8348 & 207840 & B8III & 15 & 2.755 & 0.62 &  \\
8349 & 207857 & B9mnp & 20 & 2.704 & 0.50 &  \\
8355 & 208008 & B9V & 45 & 2.743 & 0.50 &  \\
8377 & 208727 & B8V & 205 & 2.758 & 0.61 &  \\
8403 & 209419 & B5III & 20 & 2.697 & 0.48 &  \\
8418 & 209819 & B8V & 135 & 2.793 & 0.70 &  \\
8452 & 210424 & B5III & 20 & 2.723 & 0.53 &  \\
8478 & 210934 & B8III & 50 & 2.736 & 0.66 &  \\
8512 & 211838 & B8IIImnp & 65 & 2.721 & 0.68 &  \\
8554 & 212986 & B5III & 30 & 2.736 & 0.49 &  \\
8705 & 216523 & B8V & 30 & 2.761 & 0.59 &  \\
8706 & 216538 & B7III-IV & 20 & 2.706 & 0.55 &  \\
8723 & 216831 & B7III & 65 & 2.714 & 0.62 &  \\
8861 & 219749 & B9p & 65 & 2.765 & 0.66 &  \\
8873 & 219927 & B8III & 15 & 2.718 & 0.64 &  \\
8887 & 220222 & B6III & 115 & 2.708 & 0.56 &  \\
8902 & 220575 & B8III & 20 & 2.717 & 0.74 &  \\
9031 & 223640 & Ap & 30 & 2.751 & 0.56 &  \\
9086 & 224906 & B9IIImnp & 35 & 2.725 & 0.68 &  \\
9087 & 224926 & B7III-IV & 60 & 2.703 & 0.50 &  \\
9091 & 224990 & B4III & 15 & 2.706 & 0.46 &  \\
9110 & 225289 & B8mnp & 40 & 2.719 & 0.63 &  \\
 &  &  &   &  &  &  \\
Group 2 \\
38 & 829 & B2V & 0 & 2.652 & 0.20 &  \\
91 & 1976 & B5IV & 125 & 2.693 & 0.38 &  \\
155 & 3379 & B2.5IV & 55 & 2.695 & 0.26 &  \\
189 & 4142 & B5V & 160 & 2.702 & 0.41 &  \\
302 & 6300 & B3V & 100 & 2.68 & 0.33 &  \\
801 & 16908 & B3V & 90 & 2.682 & 0.32 &  \\
930 & 19268 & B5V & 25 & 2.719 & 0.43 &  \\
950 & 19736 & B4V & 50 & 2.677 & 0.36 &  \\
987 & 20365 & B3V & 120 & 2.681 & 0.34 &  \\
989 & 20418 & B5V & 260 & 2.673 & 0.39 &  \\
1005 & 20756 & B5IV & 30 & 2.718 & 0.37 &  \\
1011 & 20809 & B5V & 185 & 2.696 & 0.39 &  \\
1029 & 21071 & B7V & 50 & 2.727 & 0.44 &  \\
1034 & 21278 & B5V & 50 & 2.705 & 0.40 &  \\
1044 & 21428 & B3V & 125 & 2.686 & 0.36 &  \\
1063 & 21699 & B8IIImnp & 35 & 2.696 & 0.37 &  \\
1121 & 22920 & B9IIIsp & 30 & 2.687 & 0.43 &  \\
1153 & 23466 & B3V & 90 & 2.688 & 0.31 &  \\
1174 & 23793 & B3V & 20 & 2.684 & 0.34 &  \\
1194 & 24155 & B9sp & 35 & 2.706 & 0.34 &  \\
1199 & 24263 & B5V & 105 & 2.711 & 0.44 &  \\
1244 & 25340 & B5V & 80 & 2.708 & 0.43 &  \\
1253 & 25558 & B3V & 30 & 2.694 & 0.32 &  \\
1258 & 25631 & B2.5V & 220 & 2.654 & 0.18 &  \\
1312 & 26739 & B5IV & 30 & 2.677 & 0.40 &  \\
1350 & 27396 & B4IV & 15 & 2.678 & 0.36 &  \\
1378 & 27778 & B3V & 105 & 2.682 & 0.34 &  \\
1415 & 28355 & B3V & 0 & 2.713 & 0.40 &  \\
1553 & 30870 & B5V & 110 & 2.701 & 0.37 &  \\
1582 & 31512 & B6V & 80 & 2.695 & 0.43 &  \\
1617 & 32249 & B3V & 30 & 2.656 & 0.23 &  \\
1640 & 32612 & B2.5IV & 65 & 2.665 & 0.18 &  \\
1641 & 32630 & B3V & 95 & 2.684 & 0.31 &  \\
1719 & 34233 & B5V & 35 & 2.71 & 0.42 &  \\
1731 & 34447 & B3IV & 10 & 2.657 & 0.24 &  \\
1748 & 34748 & B1.5Vn & 295 & 2.639 & 0.14 &  \\
1749 & 34759 & B3V & 55 & 2.717 & 0.38 &  \\
1753 & 34798 & B5IV/V & 30 & 2.703 & 0.37 &  \\
1765 & 35039 & B2IV-V & 5 & 2.628 & 0.16 &  \\
1786 & 35407 & B4IVn & 295 & 2.683 & 0.32 &  \\
1798 & 35532 & B2Vn & 260 & 2.623 & 0.24 &  \\
1810 & 35708 & B2.5IV & 10 & 2.642 & 0.14 &  \\
1820 & 35912 & B2V & 5 & 2.671 & 0.20 &  \\
1839 & 36267 & B5V & 155 & 2.724 & 0.41 &  \\
1848 & 36430 & B2V & 15 & 2.674 & 0.19 &  \\
1851 & 36485 & B2V & 35 & 2.653 & 0.17 &  \\
1864 & 36653 & B3V & 155 & 2.685 & 0.31 &  \\
1871 & 36741 & B2V & 175 & 2.65 & 0.15 &  \\
1875 & 36819 & B2.5IV & 105 & 2.671 & 0.26 &  \\
1891 & 37016 & B2.5V & 100 & 2.687 & 0.25 &  \\
1898 & 37040 & B2.5IV & 145 & 2.68 & 0.24 &  \\
1900 & 37055 & B3IV & 50 & 2.716 & 0.30 &  \\
1906 & 37150 & B3Vvar & 190 & 2.653 & 0.14 &  \\
1924 & 37367 & B2IV-V & 20 & 2.629 & 0.19 &  \\
1928 & 37438 & B3IV & 40 & 2.668 & 0.25 &  \\
1946 & 37711 & B3IV & 105 & 2.672 & 0.31 &  \\
1951 & 37752 & B8p & 35 & 2.697 & 0.40 &  \\
2005 & 38804 & B5V & 40 & 2.713 & 0.43 &  \\
2128 & 40967 & B5III & 45 & 2.679 & 0.35 &  \\
2161 & 41814 & B3V & 35 & 2.718 & 0.28 &  \\
2198 & 42545 & B5Vn & 245 & 2.68 & 0.40 &  \\
2199 & 42560 & B3IV & 160 & 2.669 & 0.33 &  \\
2205 & 42690 & B2V & 0 & 2.634 & 0.20 &  \\
2213 & 42927 & B3III & 95 & 2.682 & 0.26 &  \\
2224 & 43157 & B5V & 30 & 2.673 & 0.29 &  \\
2232 & 43317 & B3IV & 130 & 2.665 & 0.30 &  \\
2266 & 43955 & B2/B3V & 40 & 2.661 & 0.24 &  \\
2273 & 44112 & B2.5V & 95 & 2.65 & 0.22 &  \\
2282 & 44402 & B2.5V & 25 & 2.677 & 0.26 &  \\
2292 & 44700 & B3V & 0 & 2.696 & 0.33 &  \\
2306 & 44953 & B8III & 30 & 2.691 & 0.31 &  \\
2325 & 45321 & B2.5V & 160 & 2.684 & 0.31 &  \\
2344 & 45546 & B2V & 70 & 2.652 & 0.20 &  \\
2361 & 45813 & B4V & 120 & 2.692 & 0.37 &  \\
2380 & 46189 & B2.5V & 85 & 2.673 & 0.30 &  \\
2433 & 47247 & B3V & 70 & 2.72 & 0.39 &  \\
2490 & 48879 & B4IV & 105 & 2.668 & 0.35 &  \\
2494 & 48977 & B2.5V & 20 & 2.669 & 0.25 &  \\
2509 & 49333 & B7II/III & 65 & 2.689 & 0.32 &  \\
2537 & 50012 & B2IV & 70 & 2.62 & 0.16 &  \\
2544 & 50093 & B2III/IV & 150 & 2.632 & 0.21 &  \\
2603 & 51630 & B2III/IV & 20 & 2.616 & 0.15 &  \\
2611 & 51823 & B2.5V & 70 & 2.676 & 0.29 &  \\
2614 & 51925 & B2.5III & 165 & 2.636 & 0.15 &  \\
2616 & 52018 & B2V & 50 & 2.666 & 0.25 &  \\
2621 & 52140 & B3V & 0 & 2.702 & 0.35 &  \\
2623 & 52273 & B2III & 80 & 2.605 & 0.16 &  \\
2625 & 52348 & B3V & 145 & 2.677 & 0.31 &  \\
2640 & 52670 & B2V & 17 & 2.635 & 0.28 &  \\
2680 & 54031 & B3V & 0 & 2.689 & 0.32 &  \\
2688 & 54224 & B2IV-V & 10 & 2.645 & 0.19 &  \\
2695 & 54669 & B4V & 115 & 2.645 & 0.15 &  \\
2704 & 54912 & B2.5IV & 40 & 2.639 & 0.19 &  \\
2718 & 55522 & B2IV/V & 95 & 2.664 & 0.26 &  \\
2756 & 56342 & B3V & 5 & 2.676 & 0.34 &  \\
2774 & 56876 & B2IV-V & 310 & 2.676 & 0.37 &  \\
2799 & 57573 & B2.5V & 190 & 2.667 & 0.33 &  \\
2800 & 57593 & B2.5V & 105 & 2.659 & 0.25 &  \\
2824 & 58325 & B2IV-V & 5 & 2.645 & 0.17 &  \\
2921 & 60855 & B2/B3V & 240 & 2.609 & 0.18 &  \\
2948 & 61555 & B6V & 107 & 2.714 & 0.28 &  \\
3192 & 67797 & B5IV & 140 & 2.684 & 0.39 &  \\
3194 & 67880 & B2.5V & 15 & 2.63 & 0.20 &  \\
3454 & 74280 & B3V & 95 & 2.653 & 0.24 &  \\
3849 & 83754 & B5V & 150 & 2.7 & 0.40 &  \\
4456 & 100600 & B4V & 140 & 2.688 & 0.32 &  \\
5191 & 120315 & B3V & 150 & 2.694 & 0.30 &  \\
5764 & 138485 & B2Vn & 210 & 2.654 & 0.17 &  \\
5902 & 142096 & B2.5V & 155 & 2.703 & 0.25 &  \\
5904 & 142114 & B2.5Vn & 250 & 2.68 & 0.22 &  \\
5907 & 142184 & B2V & 280 & 2.662 & 0.25 &  \\
5912 & 142301 & B8III/IV & 90 & 2.693 & 0.29 &  \\
5915 & 142378 & B2/B3V & 225 & 2.69 & 0.32 &  \\
5928 & 142669 & B2IV-V & 120 & 2.647 & 0.15 &  \\
5942 & 142990 & B5IV & 125 & 2.682 & 0.24 &  \\
5985 & 144218 & B2V & 65 & 2.675 & 0.14 &  \\
5988 & 144334 & B8V & 55 & 2.721 & 0.34 &  \\
5998 & 144661 & B8IV/V & 45 & 2.708 & 0.39 &  \\
6042 & 145792 & B6IV & 30 & 2.725 & 0.39 &  \\
6092 & 147394 & B5IV & 30 & 2.702 & 0.44 &  \\
6141 & 148605 & B2V & 175 & 2.662 & 0.18 &  \\
6502 & 158148 & B5V & 240 & 2.688 & 0.43 &  \\
6588 & 160762 & B3IV & 0 & 2.661 & 0.29 &  \\
6692 & 163685 & B3IV & 10 & 2.679 & 0.37 &  \\
6741 & 164900 & B3Vn & 260 & 2.67 & 0.32 &  \\
6851 & 168199 & B5V & 175 & 2.687 & 0.39 &  \\
6873 & 168797 & B3Ve & 260 & 2.644 & 0.24 &  \\
6924 & 170111 & B3V & 120 & 2.692 & 0.34 &  \\
6946 & 170740 & B2V & 25 & 2.641 & 0.15 &  \\
6984 & 171780 & B5Vne & 230 & 2.688 & 0.42 &  \\
7033 & 173087 & B5V & 85 & 2.712 & 0.37 &  \\
7081 & 174179 & B3IVpsh & 5 & 2.657 & 0.29 &  \\
7100 & 174585 & B3IV & 145 & 2.66 & 0.25 &  \\
7121 & 175191 & B2.5V & 165 & 2.667 & 0.21 &  \\
7166 & 176162 & B5IV & 200 & 2.703 & 0.35 &  \\
7179 & 176502 & B3V & 0 & 2.667 & 0.32 &  \\
7185 & 176582 & B5IV & 65 & 2.692 & 0.26 &  \\
7210 & 177003 & B2.5IV & 10 & 2.675 & 0.24 &  \\
7258 & 178329 & B3V & 0 & 2.69 & 0.32 &  \\
7306 & 180554 & B4IV & 80 & 2.677 & 0.39 &  \\
7347 & 181858 & B3IVp & 0 & 2.696 & 0.34 &  \\
7355 & 182180 & B2Vn & 320 & 2.666 & 0.23 &  \\
7372 & 182568 & B3IV & 100 & 2.667 & 0.22 &  \\
7374 & 182618 & B5V & 35 & 2.706 & 0.41 &  \\
7467 & 185330 & B5II-III & 15 & 2.667 & 0.37 &  \\
7565 & 187811 & B2.5Ve & 195 & 2.667 & 0.28 &  \\
7647 & 189687 & B3IVe & 230 & 2.638 & 0.28 &  \\
7651 & 189775 & B5III & 85 & 2.668 & 0.32 &  \\
7656 & 189944 & B4V & 10 & 2.704 & 0.44 &  \\
7700 & 191263 & B3V & 60 & 2.68 & 0.30 &  \\
7739 & 192685 & B3V & 160 & 2.658 & 0.27 &  \\
7862 & 196035 & B3IV & 20 & 2.688 & 0.28 &  \\
7870 & 196178 & B9sp & 50 & 2.709 & 0.38 &  \\
7899 & 196775 & B3V & 210 & 2.665 & 0.25 &  \\
7911 & 197018 & B6IIImnp & 55 & 2.699 & 0.44 &  \\
7912 & 197036 & B5IV & 135 & 2.679 & 0.35 &  \\
7927 & 197419 & B2IV-Ve & 105 & 2.655 & 0.28 &  \\
7929 & 197511 & B2V & 35 & 2.651 & 0.30 &  \\
7996 & 198820 & B3III & 15 & 2.677 & 0.38 &  \\
8029 & 199661 & B2.5IV & 130 & 2.682 & 0.28 &  \\
8064 & 200595 & B3Vn & 285 & 2.682 & 0.42 &  \\
8136 & 202654 & B4IV & 160 & 2.68 & 0.29 &  \\
8301 & 206672 & B3IV & 55 & 2.642 & 0.29 &  \\
8341 & 207563 & B2V & 85 & 2.647 & 0.25 &  \\
8356 & 208057 & B3Ve & 110 & 2.673 & 0.31 &  \\
8439 & 210191 & B2.5IV & 10 & 2.627 & 0.23 &  \\
8520 & 212076 & B2IV-Ve & 80 & 2.608 & 0.16 &  \\
8528 & 212222 & B5V & 35 & 2.715 & 0.43 &  \\
8535 & 212454 & B8III-IV & 40 & 2.694 & 0.42 &  \\
8549 & 212883 & B2V & 5 & 2.659 & 0.17 &  \\
8553 & 212978 & B2V & 120 & 2.646 & 0.16 &  \\
8579 & 213420 & B2IV & 70 & 2.625 & 0.17 &  \\
8758 & 217543 & B3Vpesh & 305 & 2.662 & 0.23 &  \\
8768 & 217811 & B2V & 0 & 2.661 & 0.25 &  \\
8770 & 217833 & B9IIIwe & 30 & 2.691 & 0.37 &  \\
8777 & 217943 & B2V & 145 & 2.664 & 0.18 &  \\
9011 & 223229 & B3IV & 30 & 2.68 & 0.29 &  \\
 &  &  &   &  &  &  \\
Group 3 \\
39 & 886 & B2IV & 0 & 2.629 & 0.12 &  \\
153 & 3360 & B2IV & 10 & 2.625 & 0.13 &  \\
779 & 16582 & B2IV & 5 & 2.624 & 0.10 &  \\
1072 & 21803 & B2IV & 40 & 2.61 & 0.06 &  \\
1191 & 24131 & B1V & 95 & 2.623 & 0.01 &  \\
1215 & 24640 & B1.5V & 150 & 2.646 & 0.09 &  \\
1463 & 29248 & B2III & 20 & 2.612 & 0.07 &  \\
1552 & 30836 & B2III+ & 35 & 2.606 & 0.13 &  \\
1595 & 31726 & B1V & 5 & 2.634 & 0.05 &  \\
1679 & 33328 & B2IVne & 325 & 2.604 & 0.05 &  \\
1770 & 35149 & B1V & 220 & 2.617 & 0.05 &  \\
1781 & 35299 & B1.5V & 0 & 2.636 & 0.05 &  \\
1783 & 35337 & B2IV & 15 & 2.632 & 0.06 &  \\
1790 & 35468 & B2III & 55 & 2.613 & 0.11 &  \\
1833 & 36166 & B2V & 125 & 2.641 & 0.10 &  \\
1840 & 36285 & B2IV-V & 15 & 2.646 & 0.11 &  \\
1842 & 36351 & B1.5V & 20 & 2.639 & 0.11 &  \\
1855 & 36512 & B0V & 10 & 2.597 & 0.09 &  \\
1861 & 36591 & B1IV & 5 & 2.61 & -0.01 &  \\
1873 & 36779 & B2.5V & 175 & 2.649 & 0.13 &  \\
1886 & 36959 & B1Vvar & 5 & 2.629 & 0.04 &  \\
1887 & 36960 & B0.5V & 20 & 2.6 & 0.05 &  \\
1892 & 37018 & B1V & 20 & 2.619 & -0.01 &  \\
1896 & 37023 & B0.5Vp & 80 & 2.61 & 0.00 &  \\
1911 & 37209 & B1V & 35 & 2.63 & 0.04 &  \\
1913 & 37232 & B2IV-V & 110 & 2.642 & 0.08 &  \\
1918 & 37303 & B1Vvar & 265 & 2.618 & 0.01 &  \\
1923 & 37356 & B2IV-V & 10 & 2.623 & 0.10 &  \\
1932 & 37479 & B2VpHe & 165 & 2.586 & 0.03 &  \\
1933 & 37481 & B1.5IV & 90 & 2.636 & 0.06 &  \\
1950 & 37744 & B1.5V & 25 & 2.634 & 0.04 &  \\
2031 & 39291 & B2IV-V & 150 & 2.639 & 0.10 &  \\
2058 & 39777 & B1.5V & 20 & 2.65 & 0.11 &  \\
2294 & 44743 & B1II/III & 17 & 2.593 & 0.00 &  \\
2373 & 46064 & B2III & 60 & 2.638 & 0.14 &  \\
2387 & 46328 & B1III & 0 & 2.585 & 0.02 &  \\
2618 & 52089 & B2Iab: & 25 & 2.577 & 0.00 &  \\
2628 & 52437 & B3Vnn & 190 & 2.622 & 0.14 &  \\
2648 & 52918 & B1V & 270 & 2.591 & 0.01 &  \\
2733 & 55856 & B2IV & 40 & 2.614 & 0.10 &  \\
2734 & 55857 & B0.5V & 150 & 2.605 & 0.07 &  \\
2739 & 55879 & B0III & 30 & 2.588 & 0.09 &  \\
2743 & 55985 & B2IV-V & 30 & 2.632 & 0.13 &  \\
2928 & 61068 & B2III & 10 & 2.62 & 0.05 &  \\
3004 & 62747 & B1.5III & 95 & 2.6 & 0.06 &  \\
3023 & 63271 & B2IV-V & 30 & 2.643 & 0.08 &  \\
5885 & 141637 & B1.5Vn & 225 & 2.637 & 0.10 &  \\
6453 & 157056 & B2IV & 30 & 2.623 & 0.10 &  \\
6601 & 161056 & B1.5V & 235 & 2.594 & 0.05 &  \\
6684 & 163472 & B2IV-V & 75 & 2.63 & 0.10 &  \\
6719 & 164432 & B2IV & 50 & 2.614 & 0.13 &  \\
6787 & 166182 & B2IV & 30 & 2.608 & 0.15 &  \\
7335 & 181409 & B2IVe & 140 & 2.615 & 0.02 &  \\
7940 & 197770 & B2III & 55 & 2.596 & 0.12 &  \\
8007 & 199140 & B2III & 45 & 2.611 & 0.02 &  \\
8238 & 205021 & B1IV & 20 & 2.628 & 0.01 &  \\
8603 & 214167 & B2Ve & 265 & 2.609 & 0.01 &  \\
8640 & 214993 & B2III & 30 & 2.612 & 0.04 &  \\
8651 & 215191 & B1V & 180 & 2.633 & 0.06 &  \\
8725 & 216916 & B2IV & 10 & 2.639 & 0.08 &  \\
8733 & 217101 & B2IV-V & 130 & 2.663 & 0.10 &  \\
9005 & 223128 & B2IV & 10 & 2.634 & 0.08 &  \\
9071 & 224572 & B1V & 150 & 2.608 & 0.06 &  \\
 &  &  &   &  &  &  \\
\enddata




\end{deluxetable}


\begin{deluxetable}{rrlrrrlr}
\tabletypesize{\scriptsize}



\tablecaption{Data for Field Star Binaries}

\tablenum{3}

\tablehead{\colhead{HR} & \colhead{HD} & \colhead{Sp. Type} & 
\colhead{\textit{v sin i}} & \colhead{$\beta$} & \colhead{$c_0$} 
& \colhead{K} & \colhead{} \\ 
\colhead{} & \colhead{} & \colhead{}& \colhead{(km s$^{-1}$)} &
\colhead{} & \colhead{} 
& \colhead{(km s$^{-1}$)} & \colhead{}} 

\startdata
Group 1 &  &  &  &  &  &  \\
354 &   7157 & B9V   & 22 & 2.789 & 0.632 & abt:sb2 & \\
936 &   19356 & B8V   & 50 & 2.748 & 0.616 & sb2:44 & \\
1033 & 21203 & B9V+... & 50 & 2.791 & 0.684 & abt:sb2 &\\
1088 & 22203 & B9V   & 65 & 2.765 & 0.702 & sb2:103 & \\
1185 & 23950 & B8III & 60 & 2.746 & 0.607 & abt:sb2 & \\
1240 & 25267 & B6V+... & 30 & 2.733 & 0.611 & sb1:41 & \\
1347 & 27376 & B9V   & 20 & 2.768 & 0.63 & sb2:64 & \\
1471 & 29365 & B8V   & 70 & 2.757 & 0.601 & sb1:65 &\\
1690 & 33647 & B9Vn & 30 & 2.771 & 0.687 & abt:sb2 & \\
1847 & 36408 & B7IIIe & 45 & 2.711 & 0.706 & abt:sb2 & \\
2226 & 43179 & B7V   & 25 & 2.763 & 0.616 & abt:sb2 & \\
2784 & 57103 & B8V   & 55 & 2.774 & 0.648 & sb2:106 & \\
3623 & 78316 & B8IIImnp & 15 & 2.717 & 0.558 & sb1:67 & \\
5801 & 139160 & B7IV & 25 & 2.74 & 0.451 & sb1:37 & \\
5863 & 140873 & B8III & 80 & 2.744 & 0.499 & sb2:43 & \\
5906 & 142165 & B5V   & 220 & 2.74 & 0.481 & sb1:33 & \\
6620 & 161701 & B9V   & 25 & 2.754 & 0.641 & sb1:52 & \\
6928 & 170200 & B8III-IV & 40 & 2.745 & 0.631 & sb1:38 & \\
7109 & 174853 & B8Vnn & 100 & 2.763 & 0.636 & abt:sb2 & \\
7171 & 176301 & B7III-IV & 75 & 2.707 & 0.593 & abt:sb2 & \\
7174 & 176318 & B7IV & 120 & 2.75 & 0.579 & sb1:76 & \\
7326 & 181182 & B8III+ & 60 & 2.725 & 0.603 & sb2:70 & \\
8036 & 199892 & B7III & 20 & 2.731 & 0.535 & sb1:43 & \\
8357 & 208095 & B6IV-V & 95 & 2.747 & 0.508 & sb2:106 & \\
8567 & 213236 & B8II & 15 & 2.737 & 0.669 & abt:sb2 & \\
\\
Group 2 &  &  &  &  &  &  \\
154 &   3369 & B5V &   25 & 2.688 & 0.433 & sb2:48 & \\
226 &   4727 & B5V &   20 & 2.71 & 0.407 & sb2:72 & \\
1163 & 23625 & B2.5V & 20 & 2.654 & 0.146 & sb2:82 & \\
1239 & 25204 & B3V+... & 50 & 2.691 & 0.352 & sb2:57 & \\
1288 & 26326 & B4V &   5 & 2.69 & 0.391 & abt:sb2 & \\
1420 & 28475 & B5V &   30 & 2.711 & 0.432 & sb2:80 & \\
1497 & 29763 & B3V &   115 & 2.713 & 0.356 & sb2:54 & \\
1659 & 32990 & B2V &   55 & 2.64 & 0.241 & sb1:37 & \\
1764 & 35007 & B3V &   35 & 2.691 & 0.277 & abt:sb2 & \\
1803 & 35588 & B2.5V & 170 & 2.663 & 0.189 & sb1:71 & \\
1863 & 36646 & B4Vn & 180 & 2.69 & 0.271 & abt:sb2 & \\
1890 & 37017 & B1.5V & 165 & 2.65 & 0.136 & sb1:36 & \\
2052 & 39698 & B2V &   115 & 2.635 & 0.185 & sb2:70 & \\
2159 & 41753 & B3V &   30 & 2.653 & 0.317 & sb1:33 & \\
2598 & 51411 & B3V &   0 & 2.707 & 0.376 & abt:sb2 & \\
5812 & 139365 & B2.5V & 100 & 2.689 & 0.263 & sb2:75 & \\
5934 & 142883 & B3V &   5 & 2.721 & 0.317 & sb1:64 & \\
6028 & 145482 & B2V &   165 & 2.657 & 0.177 & sb1:32 & \\
6414 & 156247 & B5Vnn & 110 & 2.695 & 0.390 & sb2:180 & \\
6431 & 156633 & B1.5V & 140 & 2.648 & 0.206 & sb2:98 & \\
6621 & 161756 & B4IVe & 95 & 2.665 & 0.349 & abt:sb2 & \\
6738 & 164852 & B3IV & 150 & 2.685 & 0.295 & sb2:58 & \\
6773 & 165814 & B4IV & 200 & 2.674 & 0.389 & sb2:151 & \\
7131 & 175426 & B2.5V & 110 & 2.666 & 0.303 & sb1:40 & \\
7305 & 180553 & B8V &   70 & 2.704 & 0.362 & sb1:55 &
\\
7486 & 185936 & B5V &   65 & 2.704 & 0.426 & sb1:47 & \\
7688 & 190993 & B3V &   115 & 2.686 & 0.293 & abt:sb2 & \\
7777 & 193536 & B2V &   80 & 2.63 & 0.219 & sb2:115 & \\
7861 & 195986 & B4III & 0 & 2.672 & 0.385 & sb1:32 & \\
8001 & 199081 & B5V &   40 & 2.713 & 0.397 & sb2:112 & \\
8384 & 208947 & B2V &   125 & 2.654 & 0.188 & sb2:116 & \\
8397 & 209288 & B5IIIn & 75 & 2.7 & 0.390 & abt:sb2 & \\
8427 & 209961 & B2V &   145 & 2.638 & 0.167 & sb1:122 & \\
8606 & 214240 & B3V &   55 & 2.678 & 0.374 & sb1:82 & \\
8800 & 218407 & B2V &   100 & 2.639 & 0.165 & sb1:86 & \\
8803 & 218440 & B2V &   25 & 2.664 & 0.166 & sb2:88 & \\
8808 & 218537 & B3V &   80 & 2.695 & 0.255 & abt:sb2 & \\
8926 & 221253 & B3IV & 140 & 2.681 & 0.295 & sb1:57 & \\
\\
Group 3 \\
1131 & 23180 & ---   & 100   & 2.597 & 0.022 & sb2;112 & \\
1333 & 27192 & B1.5IV & 110   & 2.604 & 0.027 & abt:sb2 & \\
1567 & 31237 & B3III+ & 90   & 2.603 & 0.135 & sb1:58 & \\
1811 & 35715 & B2IV & 110   & 2.622 & 0.027 & sb2:139 & \\
1868 & 36695 & B1V &   120   & 2.625 & 0.012 & sb2;129 & \\
1894 & 37021 & B0V &   240   & 2.602 & 0.125 & sb2:66 & \\
1952 & 37756 & B2IV-V & 75   & 2.632 & 0.111 & sb2:89 & \\
4590 & 104337 & B1.5V & 95   & 2.607 & 0.078 & sb2:120 & \\
5056 & 116658 & B1III-IV & 130   & 2.607 & 0.018 & sb2:120 & \\
5944 & 143018 & B1V &   90   & 2.614 & 0.016 & sb2:124 & \\
5984 & 144217 & B1V &   100   & 2.615 & -0.017 & sb2:125 & \\
7200 & 176819 & B2IV-V & 40   & 2.640 & 0.095 & sb1:55 & \\
\enddata




\end{deluxetable}


\begin{deluxetable}{lcccc}
\tabletypesize{\scriptsize}



\tablecaption{Average Values of \textit{v sin i}}

\tablenum{4}

\tablehead{\colhead{Definition of Sample	} & 
\colhead{\texttt{<}\textit{v sin i}\texttt{>}} & \colhead{No. of Stars} & \colhead{} & \colhead{} \\ 
\colhead{} & \colhead{(	km s$^{-1}$	)} & \colhead{} & \colhead{} 
& \colhead{} } 

\startdata
Group 1 &  &  \\
All Stars & & \\ 
\ \ h and $\chi$ Per Members & 183 & 72 \\
\ \ Field Stars & 92 & 259 \\
Binaries K \texttt{>} 30 km s$^{-1}$ excluded &  &  \\
\ \ h and $\chi$ Per Members & 184 & 62 \\
\ \ Field Stars & 96 & 234 \\
Binaries K \texttt{>} 30 km s$^{-1}$ only &  &  \\
\ \ h and $\chi$ Per Members & 183 & 10 \\
\ \ Field Stars & 56 & 25 \\
 \\
Group 2 &  &  \\
All Stars & & \\
\ \ h and $\chi$ Per Members & 145 & 82 \\
\ \ Field Stars & 93 & 215 \\
Binaries K \texttt{>} 30 km s$^{-1}$ excluded &  &  \\
\ \ h and $\chi$ Per Members & 156 & 73 \\
\ \ Field Stars & 95 & 177 \\
Binaries K \texttt{>} 30 km s$^{-1}$ only &  &  \\
\ \ h and $\chi$ Per Members & 44 & 9 \\
\ \ Field Stars & 84 & 38 \\
 \\
Group3 &  &  \\
All Stars & & \\
\ \ h and $\chi$ Per Members & 104 & 45 \\
\ \ Field Stars & 83 & 75 \\
Binaries K\texttt{>} 30 km s$^{-1}$ excluded &  &  \\
\ \ h and $\chi$ Per Members & 104 & 42 \\
\ \ Field Stars & 79 & 63 \\
Binaries K \texttt{>} 30 km s$^{-1}$ only &  &  \\
\ \ h and $\chi$ Per Members & 100 & 3 \\
\ \ Field Stars & 108 & 12 \\
\enddata




\end{deluxetable}


\begin{deluxetable}{lrrr}
\tabletypesize{\scriptsize}
\rotate



\tablecaption{Binary Frequency: K \texttt{>} 30 km s$^{-1}$}

\tablenum{5}

\tablehead{\colhead{Definition of Sample} & \colhead{Total No. of Stars} & \colhead{No. of Binaries} & \colhead{Binary Fraction} \\ 
\colhead{} & \colhead{} & \colhead{} & \colhead{} } 

\startdata
Group 1 & & & \\
\ \ Field & 259 & 25 & 0.1   \\
\ \ Binary Fraction Detected in Single Observation &  & & 0.06   \\
\ \ h and $\chi$ Per Detected Binaries & 71 & 9 & 0.13   \\
 \\
Group 2 &  &  &   \\
\ \ Field Stars & 215 & 38 & 0.18    \\
\ \ Binary Fraction Detected in Single Observation & & & 0.12   \\
\ \ h and $\chi$ Per Detected Binaries & 79 & 6 & 0.08    \\
 \\
Group 3 &  &  &   \\
\ \ Field Stars & 75 & 12 & 0.16   \\
\ \ Binary Fraction Detected in Single Observation & & & 0.13   \\
\ \ h and $\chi$ Per Detected Binaries & 45 & 3 & 0.07   \\
 \\
Total Sample &   &   &    \\
\ \ Field Stars & 549 & 75 & 0.14   \\
\ \ Binary Fraction Detected in Single Observation &  & & 0.09   \\
\ \ h and $\chi$ Per Detected Binaries & 195 & 18 & 0.09    \\
\enddata




\end{deluxetable}









\begin{references}
Abt, H.A. 2003, ApJ, 582, 420\\
Abt, H.A., \& Hunter, J.H., Jr. 1962, ApJ 136, 381\\
Abt., H.A., Levato, H., \& Grosso, M. 2002, ApJ 573, 359 (ALG)\\
Bernacca, P.L., \& Perinotto, M. 1974, A\&A, 33, 443\\
Bragg, A.E., \& Kenyon, S.J. 2002, AJ, 124, 3289\\
Brown, A.G.A., \& Verschueren, W. 1997, A\&A, 319, 811\\
Capilla, G., \& Fabregat, J. 2002, A\&A, 394, 479\\
Collins, G.W. \& Sonneborn, G.H. 1977, ApJS, 34.41\\
Crawford, D.L. 1978, AJ, 83, 48\\
Crawford, D.L., Glaspey, J.W., \& Perry, C.L. 1970, A. J., 75, 822\\
Drilling, J.S., \& Landolt, A.U. 2000, in Allen's Astrophysical Quantities, ed. A. N. Cox (4th ed.; New York: AIP), 381\\
Elmegreen, B. G.; Shadmehri, M. 2003, MNRAS 338, 817\\
Gies, D.R., \& Huang, W. 2003, IAU Symposium No. 215, in press\\
Heger, A., \& Langer, N. 2000, ApJ, 544, 1016\\
Hauck, B. \& Mermilliod, M. 1998, A\&A Suppl. Ser, 129, 431\\
Johns-Krull, C. M., \& Gafford, A. D. 2002, ApJ, 573, 685\\
Keller, S. C. 2004, PASA, in press (astro-ph/0405129)\\
Keller, S.C., Grebel, E.K., Miller, G.J., \& Yoss, K.M. 2001, AJ, 122, 248\\
K\"{o}nigl, A. 1991, ApJ Letters 370, 39\\
Maeder, A., Grebel, E.K., \& Mermilliod, J.-C. 1999, A\&A, 346, 459\\
Massey, P., \& Hunter, D.A. 1998, ApJ. 493, 180\\
Matt, S., \& Pudritz, R. E. 2004, ApJ, 607, L43\\
McKee, C. F., \& Tan, J. C. 2003, ApJ 585, 850\\
Meynet, G., \& Maeder, A. 2000, A\&A, 361, 101\\
Schaller, G., Schaerer, D., Meynet, G., \& Maeder, A. 1992, A\&AS, 96, 269\\
Shu, F. H.,  Najita, J., Ruden, S. P., \& Lizano, S. 1994,  ApJ, 429, 797\\
Sirianni, M., Nota, A., Leitherer, C., DeMarchi, G., \& Clampin, M. 2000, ApJ, 533, 203\\
Slesnick, C.L., Hillenbrand, L.A., \& Massey, P. 2002, ApJ 576, 880 (SHM)\\
Slettebak, A. 1968, ApJ 154, 933\\
Slettebak, A. 1985, ApJS, 59, 780\\
Slettebak, A., Collins, G.W.II, Parkinson, T.D., Boyce, P.B., \&
White, N.M. 1975, ApJS, 29, 137\\
Stahler, S.W. 1988 ApJ 332, 804\\
Stolte, A.,  Grebel, E.K., Brandner, W., \& Figer, D.F. 2002, A\&A 394, 459\\
Townsend, R.H.D., Owocki, S. P., \& Howarth, I. D. 2004, MNRAS, 350, 189\\
Vrancken, M., Lennon, D. J., Dufton, P. L., \& Lambert, D. L. 2000, 
A\&A, 358, 639\\
Wolff, S.C., \& Preston, G.W. 1978, ApJSuppl., 36, 497\\
Wolff, S. C., Edwards, S., \& Preston, G. W. 1982, ApJ 252, 322 (WEP)\\
Wolff, S.C., Strom, S. E., \& Hillenbrand, L.A. 2004, ApJ, 601, 979\\
\end{references}
\end{document}